\documentclass[]{aa}  

\usepackage{graphicx}
\usepackage[varg]{txfonts}
\usepackage{natbib}
\bibpunct{(}{)}{;}{a}{}{,} 

\begin{document}
   \title{Dark influences: imprints of dark satellites on dwarf galaxies}

   \subtitle{}

   \author{T.K. Starkenburg
          \inst{1}
          \and
          A. Helmi\inst{1}
          }

   \institute{Kapteyn Astronomical Institute, University of Groningen,
              P.O. Box 800, 9700 AV Groningen, The Netherlands\\
              \email{tjitske@astro.rug.nl}
             }

   \date{Received date / Accepted date }

  \abstract
 {In the context of the current $\Lambda$CDM cosmological model small dark matter haloes are abundant and satellites of dwarf galaxies are expected to be predominantly dark. Since low mass galaxies have smaller baryon fractions interactions with these satellites may leave particularly dramatic imprints.} 
  {We uncover the influence of the most massive of these dark satellites on disky dwarf galaxies and the possible dynamical and morphological transformations that result from these interactions.}
   {We use a suite of carefully set-up, controlled simulations of isolated dwarf galaxies. The primary dwarf galaxies have solely a stellar disk in the dark matter halo and the secundaries are completely devoid of baryons. We vary the disk mass, halo concentration, initial disk thickness and inclination of the satellite orbit.}
   {The disky dwarf galaxies are heated and disrupted due to the minor merger event, more extremely for higher satellite over disk mass ratios, and the morphology and kinematics are significantly altered. Moreover, for less concentrated haloes the minor merger can completely destroy the disk leaving a low-luminosity spheroidal-like galaxy instead.}
   {We conclude that dwarf galaxies are very much susceptible to being disturbed by dark galaxies and that even a minor merger event can significantly disrupt and alter the structure and kinematics of a dwarf galaxy. This process may be seen as a new channel for the formation of dwarf spheroidal galaxies.}

\maketitle

\section{Introduction}
The current $\Lambda$CDM cosmological standard model predicts an abundance of small dark matter haloes. A predominant fraction of these haloes (with masses $< 10^9 \ M_{\sun}$) is thought to be too small to have been able to host a galaxy due to processes such as reionization, photo-evaporation of gas and/or supernova outflows \citep{Gnedin2000, Hoeftetal2006, Kaufmannetal2007, Okamotoetal2008, GnedinTassisKravtsov09, LiDeLuciaHelmi2010, Sawalaetal2013} \citep[but see also][]{TaylorWebster2005, Warrenetal2007}. These haloes will remain almost devoid of stars and are therefore often called dark galaxies. The existence of such dark galaxies is a possible solution to the missing satellites problem \citep{Klypinetal1999,Mooreetal1999}.
 
The dark matter subhalo mass function at accretion is nearly self similar. Small differences arise because smaller haloes form earlier and therefore have a slightly lower normalization of the subhalo mass function at a fixed redshift \citep{vdBoschetal2005, vdBoschJiang2014}. A dwarf galaxy, or small dark matter halo, will thus have a spectrum of perturbers very similar to that of an \textit{L}$_\star$ or larger galaxy. 

In contrast to larger systems, smaller galaxies are very inefficient at forming stars \citep{Blantonetal2001, RobertsonKravtsov2008} therefore the stellar-to-halo mass ratio decreases towards lower halo masses \citep{Mosteretal2013, KormendyFreeman2014} and dwarf galaxy baryon fractions are well below the universal baryon fraction \citep{Gnedin2000, Hoeftetal2006, Crainetal2007}. These small galaxies thus have a subhalo mass function similar to that of larger galaxies, but have smaller baryonic components and most of their satellites can be expected to be dark \citep{Helmietal2012}.  
 
The effect of infalling satellites and mergers on the disk of a Milky Way-sized galaxy has often been studied in the past. These events lead to the heating of the disk where the amount of ``damage'' depends on the satellite-to-host mass ratio, the initial thickness of the disk, the orbit of the satellite and the presence of gas in the disk \citep{TothOstriker1992, Quinnetal1993, VelazquezWhite1999, Fontetal2001, Bensonetal2004, VH08, Hopkinsetal2008, Mosteretal2010, Kimetal2014}. Similar heating events should also take place on the dwarf galaxy scale but the results may be expected to be much more disruptive as the mass of any disk component present is much smaller than that of the heavier satellites. However this expectation has never been modelled and tested before. 

Mergers between dwarf galaxies have been discussed before \citep[e.g.][]{YozinBekki2012, Kazantzidisetal2011, Lokasetal2014}, but never in the context of dark companions,  although some recent studies have looked at the effect of significantly smaller subhaloes \citep[e.g.][]{BekkiChiba2006,ChakrabartiBlitz2009,Kannanetal2012,Loraetal2012,Wagner-Kaiseretal2014}. The frequency of mergers for dwarf galaxies has primarily been studied in the context of the Local Group. There, mergers of dwarf galaxies are prevalent before their infall into bigger, Milky Way-sized, haloes but mergers between subhaloes even occur occassionally after infall \citep{Anguloetal2009, Klimentowskietal2010, Deasonetal2014}.

Intriguingly, disks of smaller galaxies have on average a \lq\lq{}thicker\rq\rq{} morphology than their larger counterparts \citep{YoachimDalcanton2006, Sanchez-Janssenetal2010}. This is partly expected due to less efficient gas cooling in smaller haloes \citep{Kaufmannetal2007, RobertsonKravtsov2008} but may have an (additional) origin in dynamical effects. Moreover, many local field dwarf galaxies are highly irregular and seem disturbed or are experiencing a starburst but have no clearly detectable companion closeby \citep{Bergvall2012, Lellietal2014c}. This also seems to hold for seemingly isolated nearby galaxies \citep{Karachentsevetal2006,Karachentsevetal2008,Karachentsevetal2011}. Using very deep observations and detailed analysis a few small galaxies have recently been found to have companions \citep{Cannonetal2014, Nideveretal2013, MartinezDelgadoetal2012} or to be likely merging or be the remnant of a significant merger \citep{Ashleyetal2013, Amoriscoetal2014, Lokasetal2014}.

This argumentation leads us to propose that a merger of a dwarf galaxy with a dark satellite may be a additional path to form low-luminosity dwarf spheroidal galaxies in the field or near the outskirts of more massive haloes (systems perhaps akin to the Cetus and Tucana dwarf galaxies (although those are thought to be satellites of the Milky Way) or the KKR25 \citep{Makarovetal2012} and KKs3 \citep{Karachentsevetal2014} dwarf galaxies).

In this paper we perform a suite of simulations of isolated minor merger events of dwarf galaxies with a dark satellite. We vary the stellar-to-halo mass ratio of the dwarf galaxy, the initial thickness of the disk and the inclination of the satellite orbit but keep the mass ratio between the subhalo and the host at $20\%$. In addition we simulate one dwarf galaxy with properties similar to that of the Fornax dwarf spheroidal as derived from Schwarzschild models by \citet{Breddelsalldwarfs}. The model and simulation parameters are explained in Sect. \ref{Models} and the results are described in Sect. \ref{Results}. In Sect. \ref{Discussion} we present a discussion, and we conclude in Sect. \ref{Conclusions}.

\section{Models}
\label{Models}
We perform a set of controlled simulations of mergers between a dwarf galaxy and a dark satellite using the N-body/SPH code \textsc{gadget}3 \citep{SpYW01, Sp05}. The set-up of these experiments is inspired in the Aquarius simulations, a suite of 6 cosmological dark matter simulations with very high resolution \citep{Sp08}. This high resolution allows the identification of merger events on the dwarf galaxy scale. In these simulations we identified minor mergers with mass ratio $M_{\mathrm{sat}} : M_{\mathrm{host}} \ga 2:10$ for ``field'' host haloes. 

We take the orbit of the satellite to be almost completely radial with $r_{\mathrm{apo}}/r_{\mathrm{peri}} \sim 40$. At the start of the simulations the satellite is placed at a distance of $\sim 23$ kpc $h^{-1}$ from the center of the main halo with a radial velocity that is small with respect to the local circular velocity (so the  satellite is initially close to apocenter) and a small (prograde) tangential velocity. The inclination of the satellite orbit with the plane of the disk is 30 or 60 degrees, similar to other studies of disk thickening \citep{VH08, Mosteretal2010}.

The simulations described in this paper are fully collissionless (simulations including gas will be discussed in a forthcoming paper). Throughout this paper we assume a value for the Hubble constant of $H_0 = 100 h$ with $h = 0.73$ km s$^{-1}$ Mpc$^{-1}$.

\subsection{Initial conditions for the dwarf galaxy}
\subsubsection{Structure}
Our main isolated models of dwarf galaxies are similar to the simulated small galaxies in \citet{DVS08} and \citet{SDV08}. The dwarf galaxies are disky systems \citep{Mayeretal2001a, Mayeretal2006} embedded in a dark matter halo and have no gas and no bulge component initially. 
  
The dark matter halo has a \citet{Hernquist1990} profile 
\begin{equation}
\frac{\rho_0}{(r/a)(1+r/a)^3} 
\end{equation}
whose characteristic parameters are set by an ``equivalent'' NFW profile (Springel et al. 2005). More specifically, given the virial mass and concentration of a NFW halo, we set the Hernquist halo's total mass to $M_{\mathrm{vir}}$, and the scale radius $a$ is derived by requiring similar inner densities (i.e. $r_s \rho_{\mathrm{0,NFW}} = a \rho_{\mathrm{0,H}}$). For a halo of $M_{\mathrm{vir}} = 1.0 \times 10^{10}\ M_{\sun}\ h^{-1}$ and $c= 15$ this gives a scale radius $a=6.5\ \mathrm{kpc}\ h^{-1}$ and $r_{\mathrm{vir}}=35.1\ \mathrm{kpc}\ h^{-1}$. The chosen concentration is within the error bars of the expected concentration at $z = 0$ for a halo with $M_{\mathrm{vir}} = 10^{10} \ M_{\sun}\ h^{-1}$ according to the mass-concentration relations found by \citet{MaccioDuttonVandenBosch2008} and \citet{MunozCuartasetal2011}.

We also model a dwarf galaxy with properties similar to the Milky Way satellite Fornax. The mass and scale radius are consistent with the dynamical models of \citet{Breddelsalldwarfs} and the system has $M_{\mathrm{vir}}^{\mathrm{FNX}} = 4.0 \times 10^{9} \ M_{\sun}\ h^{-1}$ and $a=5.75\ \mathrm{kpc}\ h^{-1}$. Note that the concentration of the dark matter halo of this \emph{Fornax-analog} is lower ($c = 5$) than that of the other dwarf galaxy models ($c = 15$). 

As stated earlier, we assume that the stars are distributed in an exponential disk with a density profile
\begin{equation}
\label{diskdensprof}
\rho_{d} (R,z)=\frac{M_d}{4 \pi R_d^2 z_0} \ \mathrm{ exp}\left(-\frac{R}{R_d}\right) \ \mathrm{ sech}^2 \left(\frac{z}{z_0}\right) 
\end{equation}
The disk scale length is set following \citet{MoMaoWhite1998}, assuming a spin parameter for the halo of $0.033$, although our halos do not rotate. We explore a range of disk masses: $M_d=0.04 M_{\mathrm{vir}}$, $M_d=0.02 M_{\mathrm{vir}}$ and $M_d=0.008 M_{\mathrm{vir}}$. This allows us to test the dependence of the effect of the merger on the disk itself. In all our set-ups the mass of the main halo and the mass of the satellite are kept constant.

All our disks are required to be equally stable with similar Toomre-stability parameter $Q$ 
\begin{equation}
Q (R) = \frac{\sigma_R (R) \kappa (R)}{3.36 G \Sigma (R)}
\end{equation}
\citep{Toomre1964}, where $G$ is the gravitational constant, $\Sigma (R)$ is the surface density of the disk,  $\kappa (R)$ the epicyclic frequency and $\sigma_R (R)$ the radial velocity dispersion, all at radius $R$. Note that the influence of the mass of the disk on the epicyclic frequency $\kappa (R)$ is negligible because the halo dominates at all radii for these systems. This implies that the Toomre $Q$ varies with disk mass only through the surface density $\Sigma (R)$ and the radial velocity dispersion $\sigma_R (R)$.

We also explore disks with different initial scale heights $z_0 = 0.1 R_d$ and $z_0 = 0.2 R_d$. The Fornax-like dwarf galaxy has initial scale heights of $z_0 = 0.1 R_d$ (as in the ``standard'' set-up) and $z_0 = 0.3 R_d$, as smaller systems are expected to be thicker due to less efficient gas cooling \citep{Kaufmannetal2007,  RobertsonKravtsov2008}.

All parameters governing the structure of the systems are summarized in Table \ref{parm}. The numerical parameters of our simulations are described in more detail in Sect. \ref{numericalparameters}.

\subsubsection{Velocity structure}
\label{velstruct}
The initial conditions for the dwarf galaxies are generated following \citet{SDMH05} and \citet{Hernquist1993}. In this method the moments of the velocity distribution are calculated assuming that the distribution function only depends on the energy $E$ and the $L_z$ component of the angular momentum. Important differences arise in our set-up because the potential is dominated by the halo at all radii, and the velocity dispersions are not negligible compared to the circular velocity in a significant part of the disk. Hence the epicyclic approximation typically used breaks down not only in the very center but over a much larger extent of the disk. 

Moreover, setting up velocities for the halo particles following a Gaussian distribution with dispersions derived from the Jeans equations leads to a configuration that is not in equilibrium in the central parts of the halo \citep{KMM04, SDMH05}. This is negligible when a disk dominates the region under consideration but significant in the case of dwarf galaxies with their considerable lower disk-to-total mass ratios. Therefore we determine the velocities and dispersions for the halo particles using the distribution function for a Hernquist profile neglecting the effect of the disk. 

For the set-up of the velocity structure of the disk we start from the Jeans equations
\begin{equation}
\frac{\partial (\rho \overline{v_R^2})}{\partial R} +  \frac{\partial (\rho \overline{v_R v_z})}{\partial z} + \rho \left(\frac{\overline{v_R^2}-\overline{v_{\phi}^2}}{R} + \frac{\partial \Phi}{\partial R}\right) = 0, \label{Jeans1}
\end{equation}
\begin{equation}
\frac{1}{R} \frac{\partial (R \rho \overline{v_R v_z})}{\partial R} +  \frac{\partial (\rho \overline{v_z^2})}{\partial z} + \rho  \frac{\partial \Phi}{\partial z} = 0, \label{Jeans2}
\end{equation}
\begin{equation}
\frac{1}{R^2} \frac{\partial (R^2 \rho \overline{v_R v_{\phi}})}{\partial R} +  \frac{\partial (\rho \overline{v_z v_{\phi}})}{\partial z} = 0. \label{Jeans3}
\end{equation}
Because the halo dominates the potential at all radii we now assume that the velocity ellipsiod is aligned with the coordinate directions of the spherical coordinate system $(r, \theta, \phi)$ instead of the cylindrical system $(R, \theta, z)$ which is the usual assumption for massive disks. In our case we can approximate,
\begin{equation}
\overline{v_R v_z} \simeq \left(\overline{v_R^2} - \overline{v_z^2}\right)\left(z/R\right)
\end{equation}
\citep{BT}. In the disk ($z \sim 0$) Eq. (\ref{Jeans2}) gives the vertical velocity dispersion
\begin{equation}
\label{eqvz}
 \rho \overline{v_z^2} = \int_z^{\infty} \rho\left(R,z'\right) \frac{\partial \Phi}{\partial z'} \ \mathrm{ d}z'. 
\end{equation}
  
The vertical velocity dispersion in the midplane is related to the disk surface density. For an isothermal sheet $\overline{v_z^2} = z_0 \pi G \Sigma\left(R\right)$ \citep{Hernquist1993}, which would result in an exponential variation with radius. It is often assumed that this dependence with radius holds for both $\sigma_z$ and $\sigma_R$, i.e. that the ratio $\sigma_z/\sigma_R$ is constant, and that the epicylic frequency holds for all disks \citep{BinneyMerrifield, VanderKruitFreeman2011ARA&A, Hernquist1993, GerssenKuijkenMerrifield1997, Westfalletal2011}. This is also typically used for setting up a stable stellar disk for N-body simulations \citep{Hernquist1993, SDMH05}. However, to make the Toomre stability parameter $Q(R)$ be similar for the different disks we relax these assumptions and require that $\overline{v_R^2} = f \overline{v_z^2}$ with $f \geq 1$ constant. Using the Jeans equations to calculate the vertical velocity dispersion (see Eq. (\ref{eqvz})) we can find the radial velocity dispersion and the proportionality factor $f$ by requiring that $Q(R) \geq 2$ throughout the disk. The value $Q(R) \geq 2$ is to avoid spontaneous bar formation following \citet{AthanassoulaSellwood1986} and \citet{Athanassoula2003}. 

   \begin{figure}
   \includegraphics[width=.45\textwidth]{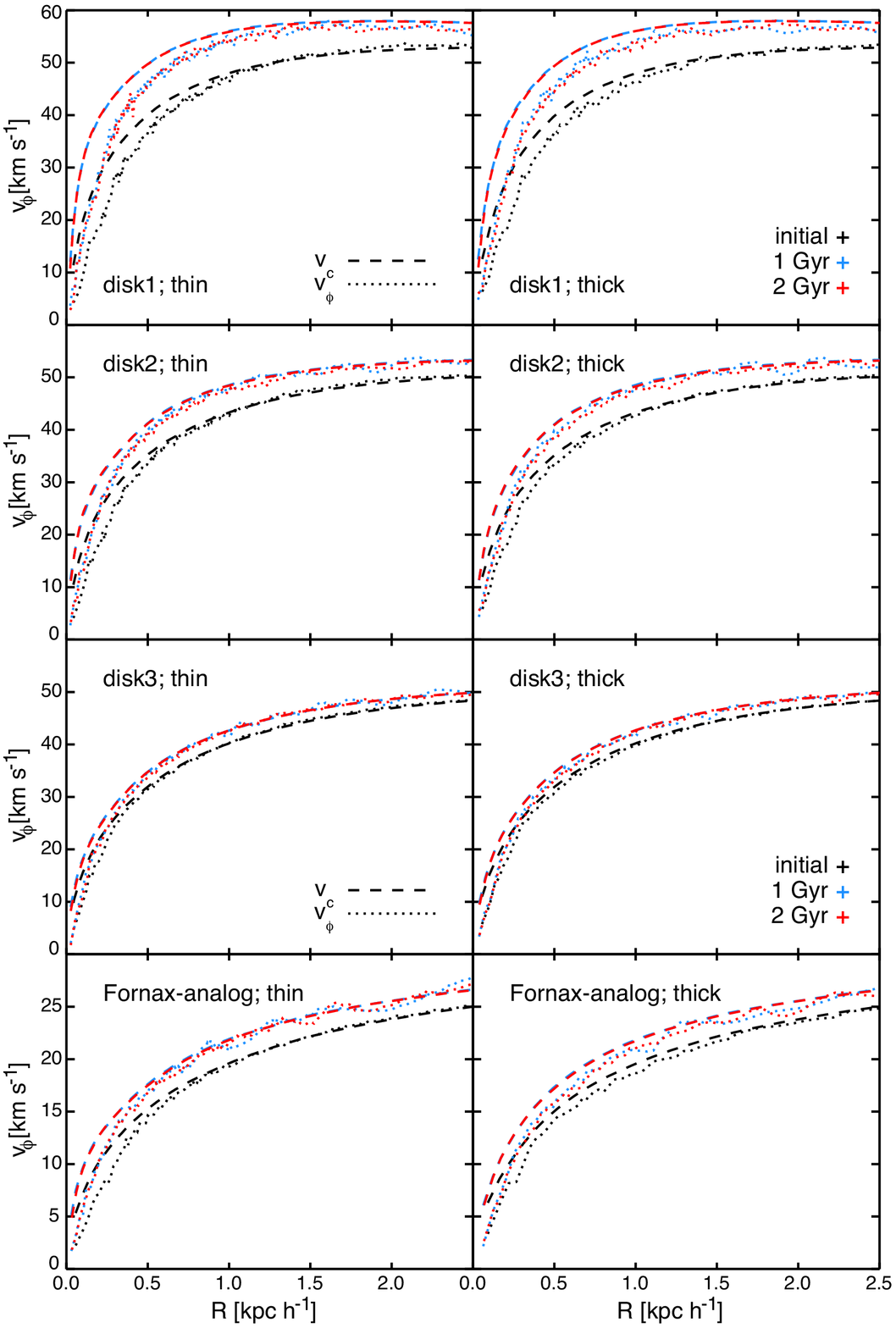}
   \caption{\label{relaxvphi} The circular (dashed lines) and azimuthal (dotted lines) velocities of all the disks when relaxed in isolation: the initial conditions (black), after 1 Gyr (blue) and after 2 Gyr (red). }
   \end{figure}

   \begin{figure}
   \includegraphics[width=.45\textwidth]{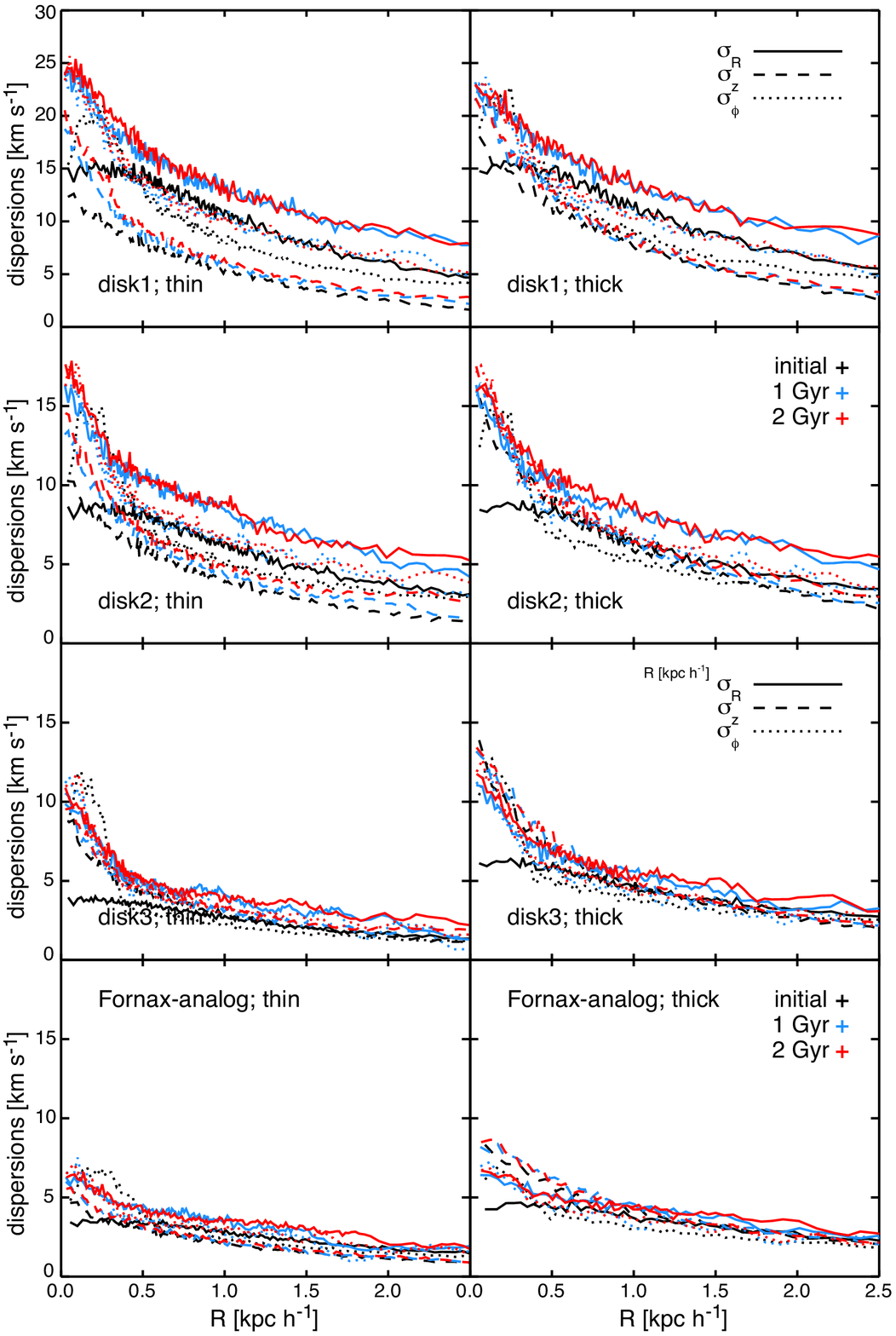}
   \caption{\label{relaxvdisp} Radial (solid lines), vertical (dashed lines) and azimuthal (dotted lines) velocity dispersions of all the disks when relaxed in isolation: the initial conditions (black), after 1 Gyr (blue) and after 2 Gyr (red).}
   \end{figure}

The velocity dispersions obtained in this way can have a very steep, close to exponential rise in the inner parts of the disk. This subsequently leads to extremely high radial velocity dispersions because we set  $\sigma_R^2/\sigma_z^2 = f$ wih $ f \geq 1$. As a consequence this can result in an imaginary azimuthal streaming velocity \citep{Quinnetal1993, Hernquist1993} if the epicyclic approximation is used. We follow \citet{Hernquist1993} in solving this by smoothing the velocity dispersions using $R'=\sqrt{R^2+a_s^2}$, where $a_s$ is the smoothing length which we set equal to half the disk scale length, $a_s = 0.5 R_d$. However, unlike \citet{Hernquist1993}, we smooth only the radial velocity dispersion as the vertical velocity dispersion is determined directly from the Jeans equations (see Eq. (\ref{eqvz})).\footnote{Note that this means that the radial and vertical velocity dispersions no longer have the same ratio throughout the disk but that the proportionality factor $f$ changes with radius, $f = f_R = \sigma_R/\sigma_z$, for the inner part of the disk.} 

We find the azimuthal velocities using the assumptions stated above and using Eq. (\ref{Jeans1}),
\begin{equation}
 \overline{v_{\phi}^2} = \frac{R}{\rho} \frac{\partial (\rho \overline{v_R^2})}{\partial R} + \overline{v_R^2} + v_c^2 - z (f_R - 1)\frac{\partial \Phi}{\partial z} + (f_R - 1)\overline{v_z^2}.
\end{equation}
Lastly, we need to find $\overline{v_{\phi}}^2$ and $\sigma^2_{\phi}$ which are related to $\overline{v_{\phi}^2}$ through
\begin{equation}
\overline{v_{\phi}}^2 = \overline{v_{\phi}^2} - \sigma^2_{\phi} \label{azvel}.
\end{equation}
To constrain these velocities normally the epicyclic approximation is used
\begin{equation}
\overline{(v_{\phi} - v_c)^2} = \frac{\overline{v_R^2}}{\eta^2},
\end{equation} where
\begin{equation}
 \eta^2 = \frac{4}{R} \frac{\partial\Phi}{\partial R} \left(\frac{3}{R}\frac{\partial\Phi}{\partial R}+\frac{\partial^2 \Phi}{\partial R^2}\right)^{-1}.
\end{equation}

This approximation, however, is only valid when the rotational motion dominates and the velocity dispersions are small. Therefore we use the epicyclic approximation for the outer parts of the disk from the radius where $\overline{v_{\phi}^2} = v_c^2$. Since $\overline{v_{\phi}}^2 \leq v_c^2$ must hold, but $\overline{v_{\phi}^2}$ is large at the center, $\sigma_{\phi}^2$ is likely to be significant there. Inwards of this point we fit quadratic functions to $v_{\phi}$ and require $v_{\phi} (R=0, z) = 0$ at each $z$. Note that although the function itself is continuous we do not force continuity on the slope of $v_{\phi}$. In that region $\sigma_{\phi}$ is then found using Eq. (\ref{azvel}).

\subsection{Satellites}
The dark satellite has a NFW profile \citep{NFW}, set up following \citet{VH08}, using the mass-concentration relation found by \citet{MunozCuartasetal2011}. The satellite has a virial mass that is 20\% of that of the host, $M_{\mathrm{sat}}=2 \times 10^9 \ M_{\sun}\ h^{-1}$, a concentration of $c = 17.25$, and has no baryonic matter at all. For the \emph{Fornax-analog} experiment the satellite has a virial mass of $0.2 M^{\mathrm{\emph{FNX}}}_{\mathrm{vir}} = 8 \times 10^8 \ M_{\sun}\ h^{-1}$ and a concentration of $c = 18.91$ (see Table \ref{parm}). We compare the average density of the host, including both the halo and disk, inside the pericenter radius with the average density of the satellite within $r_{s \mathrm{,sat}}$. The ratio $\overline{\rho_{r_s\mathrm{,sat}}}/\overline{\rho_{r_p \mathrm{,host}}}$ ranges from $0.075$ for \emph{disk1}, $0.10$ for \emph{disk2} and $0.13$ for \emph{disk3}, all with the more concentrated halo, to $0.63$ for the \emph{FNX-analog}, with the less concentrated halo host halo.

  \begin{table*}[\textwidth]
      \caption{\label{parm} Initial values for all parameters}
         $$ 
         \begin{array}{lllll}
            \hline
            \noalign{\smallskip}
              \textbf{Parameters}    & \textbf{\emph{System 1}} \qquad \qquad & \textbf{\emph{System 2}} \qquad \qquad & \textbf{\emph{System 3}} \qquad \qquad & \textbf{\emph{Fornax-analog}}   \\
            \noalign{\smallskip}
            \hline
            \hline
            \noalign{\smallskip}
            \textbf{Primary: halo}      &                &                     &                         \\
            \noalign{\smallskip}
            \hline
            \noalign{\smallskip}
            M_{\mathrm{vir}} \quad [M_{\sun}\ h^{-1}] & 1 \times 10^{10} & 1 \times 10^{10} & 1 \times 10^{10} & 4 \times 10^{9} \\
            r_{\mathrm{vir}} \quad [\mathrm{kpc}\ h^{-1}] & 35.1 & 35.1 & 35.1 & 20.2 \\
            \textrm{Hernquist scale radius } a \quad [\mathrm{kpc}\ h^{-1}] & 6.5 & 6.5 & 6.5 & 5.75 \\
            \textrm{Concentration } c    & 15 & 15 & 15 & 5 \\
            \textrm{Number of particles} & 1 \times 10^6 & 1 \times 10^6 & 5 \times 10^6 & 5 \times 10^6 \\
            \textrm{Softening length} \quad [\mathrm{pc}\ h^{-1}] & 18 & 18 & 15 & 20 \\
            \hline
            \hline
            \noalign{\smallskip}
            \textbf{Primary: disk}     &                 &                      &                       \\
            \noalign{\smallskip}
            \hline
            \noalign{\smallskip}
            M_{\mathrm{disk}} \quad [M_{\sun}\ h^{-1}] & 4 \times 10^{8} & 2 \times 10^{8} & 8 \times 10^{7} & 3.2 \times 10^{7} \\
            \textrm{Disk scale length } R_d \quad [\mathrm{kpc}\ h^{-1}] & 0.566 & 0.566 & 0.566 & 0.688 \\
            \textrm{Disk scale height } z_0  \textrm{ thin disk} \quad [\mathrm{kpc}\ h^{-1}] \qquad  & 0.1 R_d & 0.1 R_d & 0.1 R_d & 0.1 R_d \\
            \textrm{Disk scale height } z_0  \textrm{ thick disk} \quad [\mathrm{kpc}\ h^{-1}] \qquad  & 0.2 R_d & 0.2 R_d & 0.2 R_d & 0.3 R_d \\
            \textrm{Radial dispersion factor } f \textrm{ thin disk} & 7.37 & 3.84 & 1.29 & 1.29 \\
            \textrm{Radial dispersion factor } f \textrm{ thick disk} & 2.77 & 1.44 & 1.0 & 1.0 \\
            \textrm{Number of particles} & 1 \times 10^5 & 1 \times 10^5 & 1 \times 10^5 & 1 \times 10^5 \\
            \textrm{Softening length} \quad [\mathrm{pc}\ h^{-1}] & 6 & 6 & 6 & 8 \\
            \hline
            \hline
            \noalign{\smallskip}
            \textbf{Secondary: halo}      &           &            &            &          \\
            \noalign{\smallskip}
            \hline
            \noalign{\smallskip}
            M_{\mathrm{vir}} \quad [M_{\sun}\ h^{-1}] & 2 \times 10^{9} & 2 \times 10^{9} & 2 \times 10^{9} & 8 \times 10^{8} \\
            r_{\mathrm{vir}} \quad [\mathrm{kpc}\ h^{-1}] & 26.5 & 26.5 & 26.5 & 19.7 \\
            \textrm{Concentration } c & 17.25 & 17.25 & 17.25 & 18.91 \\
            \textrm{Number of particles} & 5 \times 10^5 & 5 \times 10^5 & 5 \times 10^5 & 5 \times 10^5 \\
            \textrm{Softening length} \quad [\mathrm{pc}\ h^{-1}] & 12 & 12 & 12 & 10 \\
            \hline
         \end{array}
     $$ 
   \end{table*}

\subsection{Numerical parameters}
\label{numericalparameters}
The N-body systems are evolved using the code \textsc{gadget}3. The number of particles used for the different components are chosen such that artificial bar formation due to graininess of the potential can be avoided and such that the ratio between the baryonic and dark matter particle masses is larger than $1 : 10$ respectively, to avoid numerical heating of the disk due to the dark matter halo particles. To ensure this, the systems with smaller disk masses, \emph{disk3} and the \emph{Fornax-analog}, have more particles in the host dark matter halo and a slightly different softening length.

To set the softening lengths we follow \citet{VH08} and use the prescription presented in \citet{Athanassoulaetal2000}. These authors found correlations between the optimal softening length and the mean distance from each particle to its sixth closest neighbour, 
\begin{equation}
r_{6 \mathrm{, mean}} = \left( N^{-1} \sum^N_{i=1} r_{6,i}^{-1} \right)^{-1} 
\end{equation}
for a variety of mass distributions. We determined the optimal softening for our simulations calculating $r_{6 \mathrm{, mean}}$ for our simulations after using a first guess for the softening and then comparing these values to those found by \citet{VH08}. As our central densities and mass distributions of the individual components are very similar we find $\epsilon_{\mathrm{opt}}$ using their relation between $\epsilon_{\mathrm{opt}}$ and $r_{6 \mathrm{, mean}}$. We have also checked that the derived softening lengths gave stable disk and halo profiles compared to taking slightly different values.

The criterion we use for the timestep is the standard \textsc{gadget} criterion, $\Delta t = \sqrt{2 \eta \epsilon / |a|}$, where we set  $\eta= 0.025$ and fix the maximum timestep to 25 Myr. The total energy is conserved to within $\sim 1$ percent over 6 Gyr in all simulations. 

\subsection{Evolution in isolation}

Both the host and satellite are relaxed in isolation to ensure that they are in equilibrium before the merger. Figures \ref{relaxvphi} and \ref{relaxvdisp} show the velocity structure of the host disks. The rotation and velocity dispersions are measured in cylindrical bins of variable radial binsize which contain a fixed number of 400 particles, and close to the midplane of the disk, i.e. $|z| < 0.05$ kpc $h^{-1}$.

From Fig. \ref{relaxvphi} we note that the disks rotational motions do not evolve significantly. There is a slight increase in $v_c$ and $v_{\phi}$ due to contraction of the halo but this stabilizes very quickly. Figure \ref{relaxvdisp} shows the velocity dispersions of the disks. It can be seen that our method to set the velocity distribution without relying on the epicyclic approximation near the center of the disks results in a small unrealistic bump in $\sigma_{\phi}$ at the initial time. This is caused by the approximation of the mean streaming velocity by a quadratic function where $\overline{v_{\phi}^2} \leq v_c^2$. The feature however dissappears quickly as the disk relaxes towards its equilibrium configuration.

\section{Results}
\label{Results}
\begin{figure}
\includegraphics[width=.45\textwidth]{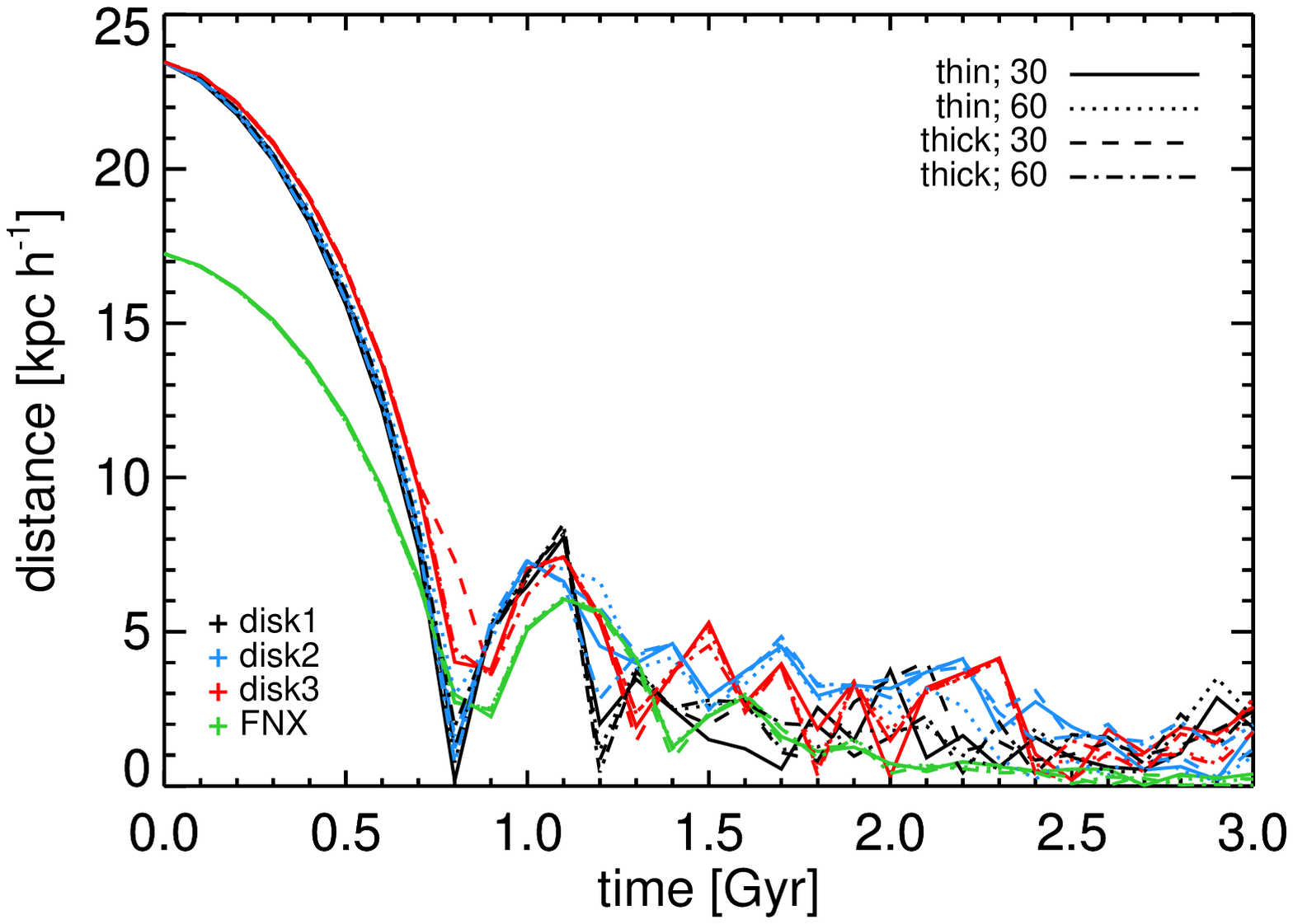}
\caption{\label{figorbits} The distance between the center of the dwarf galaxy and its satellite with time for all our simulations: the most massive disk (\emph{disk1}; black), the intermediate disk (\emph{disk2}; blue), the least massive disk (\emph{disk3}; red) and the \emph{Fornax-analog} (green) for two different inclinations of the satellite orbit with respect to the disk and two sets of initial scale heights as indicated by the top right inset. The orbits show the center of mass trajectory defined by the 100 most bound particles in the satellite. The satellite becomes completely unbound within 2 Gyr.}
\end{figure}

\begin{figure}
\includegraphics[width=.45\textwidth]{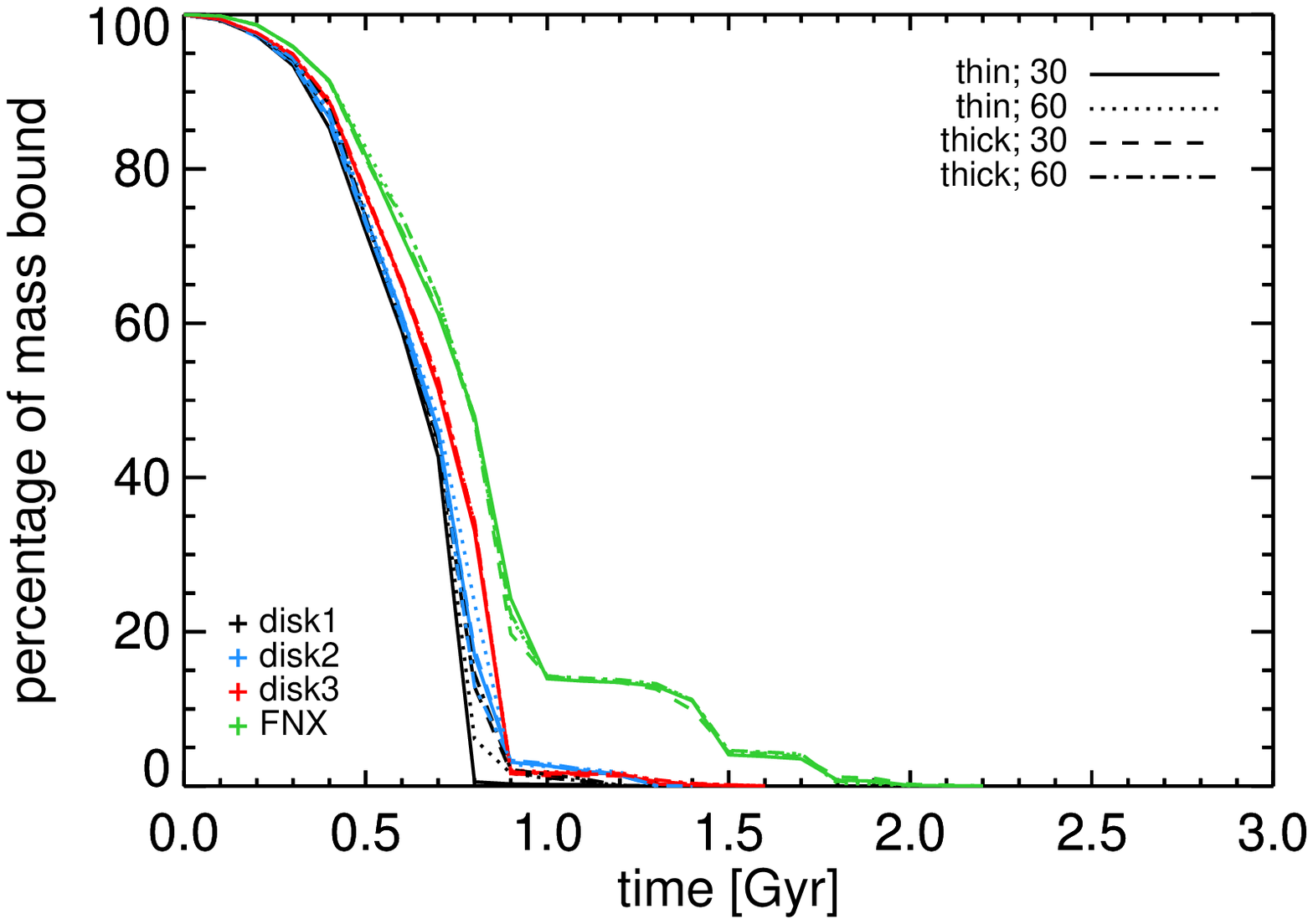}
\caption{\label{figmass} The percentage of mass still bound to the satellite as a function of time for all simulations consisting of four different dwarfs with each two different initial scale heights and two different orbits for the satellite: the most massive disk (\emph{disk1}; black), the intermediate disk (\emph{disk2}; blue), the least massive disk (\emph{disk3}; red) and the \emph{Fornax-analog} (green). The color and line coding is the same as in Fig. \ref{figorbits}}
\end{figure}

Figure \ref{figorbits} shows the orbits of the satellite for the different dwarf galaxies. At each snapshot we determine the center of the satellite by computing the center of mass using its 100 most bound particles. The satellite sinks in very quickly through dynamical friction to the central regions of the host where it is fully disrupted in a few passages. This is shown in Fig. \ref{figmass} where we have plotted the total mass that is still bound to the satellite as a function of time. Note that for all three systems with the more massive halo ($M_{\mathrm{vir}} = 1 \times 10^{10} \ M_{\sun}\ h^{-1}$) the satellite becomes almost completely unbound before second pericenter. The satellite is more severly stripped when falling into the primary with a more massive disk although the differences are marginal. This is reflected  in Fig. \ref{figmass} in the red lines lying above the blue lines which lie above the black lines.  This is an indication that the disk mass influences slightly the (minor) merger process.

\subsection{Morphological changes}
\label{morphology}

\begin{figure*}
 \centering
\includegraphics[width=0.95\textwidth]{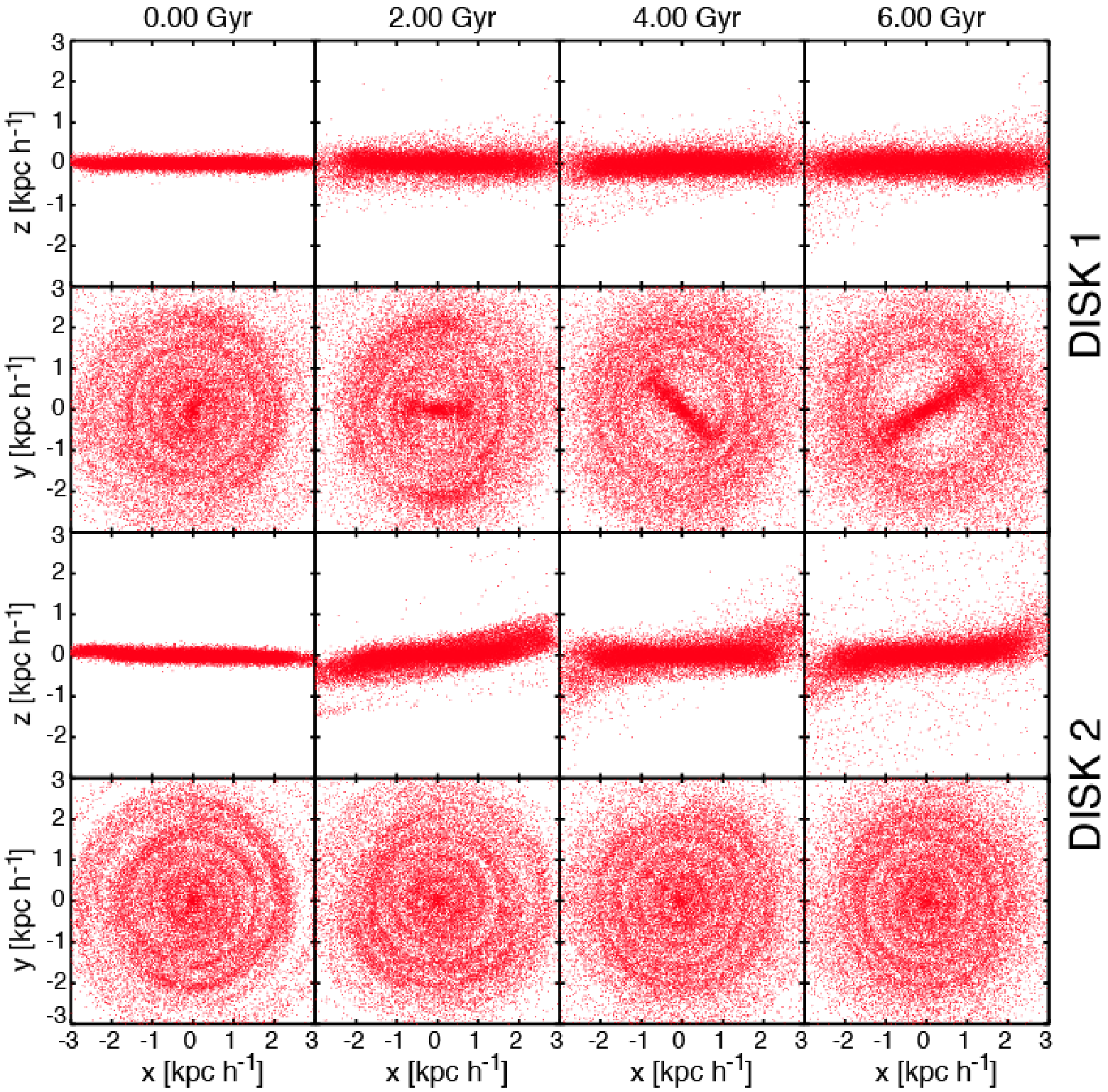}
\caption{\label{figmorph1} Edge-on and face-on particle distributions of the two most massive disks before the merger (left) upto around 5 Gyr after the first pericenter of the satellite (6 Gyr; right) in steps of 2 Gyr. The orbit of the satellite has an inclination of 60 degrees. The top two panels are for the most massive disk (\emph{disk1}) while the bottom two for the intermediate disk (\emph{disk2}), face-on and edge-on respectively, both with $z_0 = 0.1 R_d$ initially. To show the features present better we plot: within 0.5 kpc $h^{-1}$ 5\% of the particles, 10\% for radii between 0.5 and 1.0 kpc $h^{-1}$, 20\% with radii between 1.0 and 1.5 kpc $h^{-1}$ are shown, 50\% of the particles with radii between 1.5 ans 2.0 kpc $h^{-1}$ and every particle is plotted for larger radii.}
\end{figure*}

\begin{figure*}
\centering
\includegraphics[width=.95\textwidth]{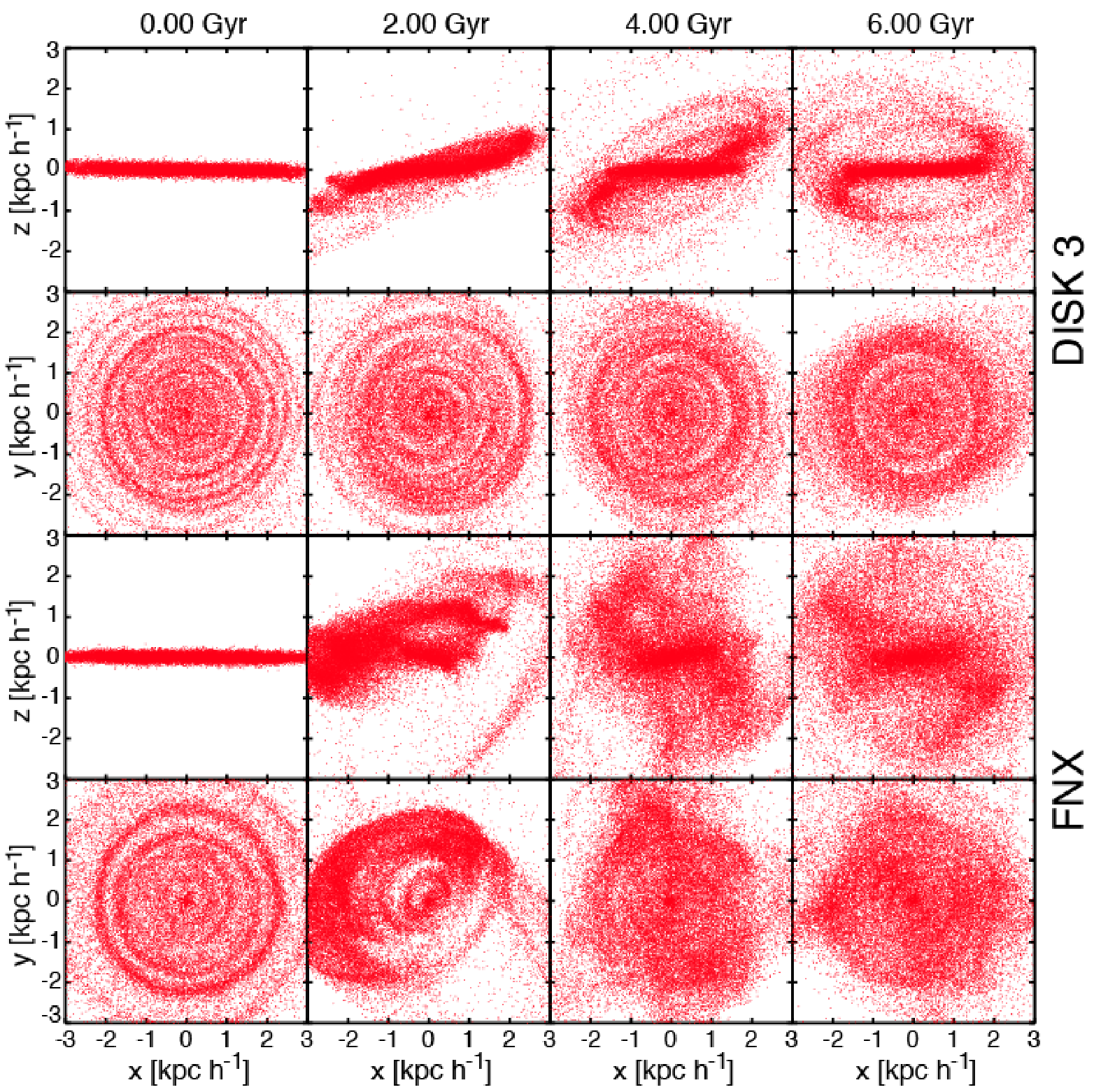}
\caption{\label{figmorph2} Edge-on and face-on particle distributions of the two least massive disks before the merger (left) upto around 5 Gyr after the first pericenter of the satellite (6 Gyr; right) in steps of 2 Gyr. The orbit of the satellite has an inclination of 60 degrees. The top two panels are for the least massive disk (\emph{disk3}) while the bottom two for the \emph{FNX-analog}, face-on and edge-on respectively, both with $z_0 = 0.1 R_d$ initially. To show the features present better we plot: within 0.5 kpc $h^{-1}$ 5\% of the particles, 10\% for radii between 0.5 and 1.0 kpc $h^{-1}$, 20\% with radii between 1.0 and 1.5 kpc $h^{-1}$ are shown, 50\% of the particles with radii between 1.5 ans 2.0 kpc $h^{-1}$ and every particle is plotted for larger radii.}
\end{figure*}

\subsubsection{Substructure}

The morphology of all the dwarf galaxies changes because of the merger as can be seen in Fig. \ref{figmorph1} (for the two most massive disks) and Fig. \ref{figmorph2} (for the least massive disk and the \emph{Fornax-analog}). In these figures we show the thinner disks, both face-on and edge-on, before, during and after the merger in steps of 2 Gyr for the satellite on the orbit with an inclination of 60 degrees. In general all the disks tilt during the merger. For all plots in this paper however, the galaxies are oriented in such a way that the direction of total angular momentum of all disk particles within the half mass radius is parallel to the $z$-axis. 

An ``envelope'' of stars coming from the disk itself can be found around the disks in these plots. The smaller mass disks end up with larger and more densely populated envelopes than the larger mass disks. All the disks have experienced thickening (see also Sects. \ref{sectshapes} and \ref{sectprof}) while largely retaining the disky-like appearance except for the \emph{FNX-analog}. \emph{Disk1} develops irregular spiral arms and a strong bar during the merger. The least massive disk on the other hand, is more clearly disturbed by the merger and develops spiral arms and tilts very strongly. A comparison of the most and least massive disks shown in the topmost panels of Figs. \ref{figmorph1} and \ref{figmorph2} gives an indication of the very different effects of the satellite on a disk of $4 \times 10^8 \ M_{\sun}\ h^{-1}$ and a disk of $8 \times 10^7 \ M_{\sun}\ h^{-1}$ in dark matter halos of exactly the same mass. 

As just mentioned, as a consequence of the merger the \emph{Fornax-analog} system cannot be called a disky dwarf galaxy from after the second pericenter of the satellite. From an initially stable disk the galaxy first becomes lopsided with warps and irregular structure in the disk and subsequently gets completely disturbed and finally more closely resembles a spheroidal system as shown in the left-most plots of the last two rows of Fig. \ref{figmorph2}.

The later snapshots of \emph{disk3} and the \emph{Fornax-analog} in Fig. \ref{figmorph2} show an abundance of tidal debris that is distributed in shells and plumes. This debris comes from the disk itself as the satellite does not have any stars. Some of the shell structures that are formed during the merger are very similar to structures observed around external galaxies and thought to be formed by secondary galaxies falling in. These simulations show that these structures can also be from the primary object itself if the merger event is significant enough.

\begin{figure*}
\centering
\includegraphics[width=\textwidth]{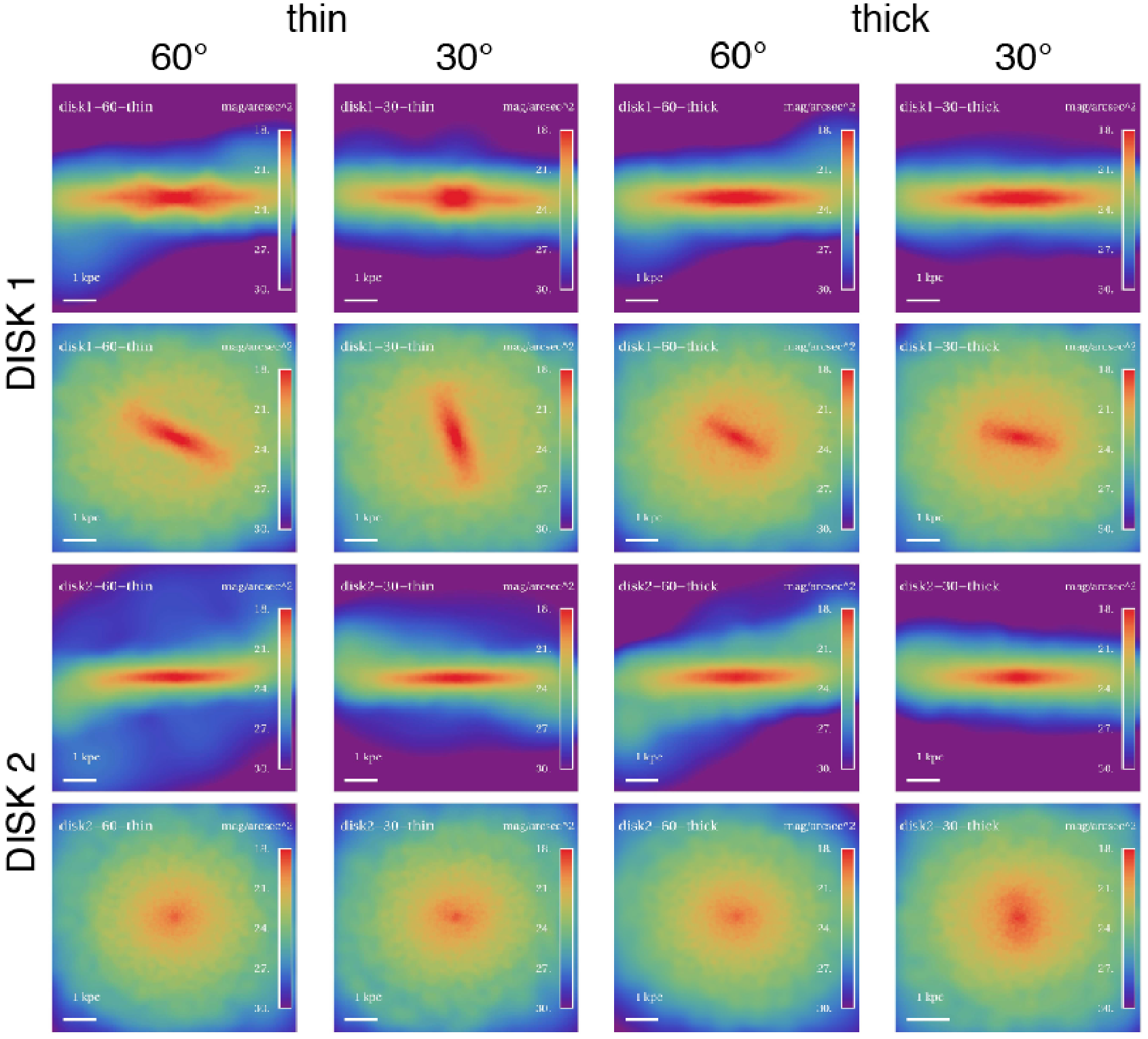}
\caption{\label{SB1} Face-on and edge-on surface brightness maps of the final snapshots for the simulations with the two most massive disks (top: \emph{disk1}, bottom: \emph{disk2}). The first two columns are for $z_0 = 0.1 R_d$, while those on the right have  $z_0 = 0.2 R_d$ initially. The first and third columns are for the 60-degrees inclination satellite, the second and fourth for the 30-degrees.}
 \end{figure*}

\begin{figure*}
\centering
\includegraphics[width=\textwidth]{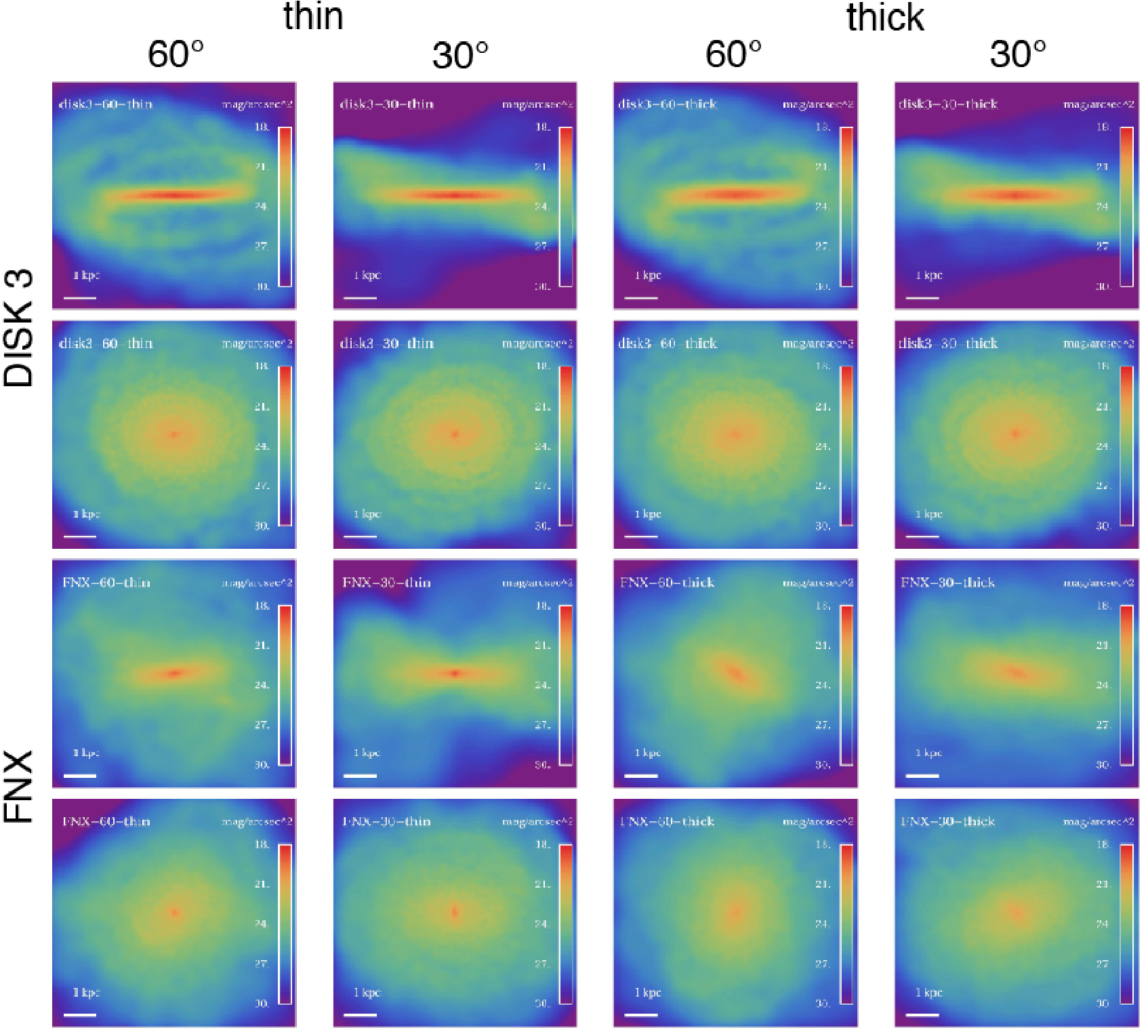}
\caption{\label{SB2} Face-on and edge-on surface brightness maps of the final snapshots for the simulations with the two least massive disks (top: \emph{disk3}, bottom: \emph{Fornax-analog}). The first two columns are for $z_0 = 0.1 R_d$, while those on the right have $z_0 = 0.2 R_d$ for \emph{disk3} and $z_0 = 0.3 R_d$ for the \emph{FNX-analog} initially. The first and third columns are for the 60-degrees inclination satellite, the second and fourth for the 30-degrees.}
 \end{figure*}

Figures \ref{SB1} and \ref{SB2} show the surface brightness of the disks, face-on and edge-on. Since magnitudes are given on a logarithmic scale the density changes are much more clearly visible than in Figs. \ref{figmorph1} and \ref{figmorph2}. Typically particle plots emphasize the small scale structures that emerge during the interaction but fail to show the density contrasts closer to the center of each galaxy. On the other hand the surface brightness maps are smoothed maps of the density distribution and so the small scale structures in the outskirts are mostly lost. However the density contrasts in the galaxies themselves are very well represented in the surface brightness maps. We compute densities by smoothing over 128 neighbours and assume a $M/L$-ratio of 2 for all star particles in all simulated galaxies. 

It is instructive to compare the rightmost columns of Figs. \ref{figmorph1} and \ref{figmorph2} with the first columns of Figs. \ref{SB1} and \ref{SB2} respectively. Figures \ref{SB1} and \ref{SB2} show surface brightness maps of the final snapshots of each simulation, of which the leftmost column corresponds to the simulations with initially thin disks and the satellite on the 60-degrees orbit.  The density gradients in the central parts of all systems as well as for example the bars developed in all the \emph{disk1}s and the spiral structures and the warps in all the \emph{disk3}s, are evident in Figs. \ref{SB1} and \ref{SB2}. Figure \ref{SB2} also shows that the \emph{FNX-60-thin} remnant is a still somewhat oblate system but that this is not true for the same set-up with an initially thicker disk (\emph{FNX-60-thick}).

The resulting surface brightness maps in Figs. \ref{SB1} and \ref{SB2} can in principle be compared to observations. Photometric images currently go down to a surface brightness of approximately $26$--$27 \ \mathrm{mag}/\mathrm{arcsec}^2$ \citep[see for example][]{Colemanetal2004, MartinezDelgadoetal2012} but occasionally go deeper as for example \citet{Hunteretal2011} who use azimuthally averaged surface brightness profiles to reach a surface brightness of $\mu_V = 28$--$31 \ \mathrm{ mag}/\mathrm{arcsec}^2$. 

The bars in all the most massive disks should all be apparent in photometric observations reaching such depths. Also some of the debris around the disks, for example for \emph{disk3-60-thin} or \emph{disk3-60-thick}, and some subtler substructure or spiral patterns in the disks would probably be visible. On the other hand most of the envelopes around the disks that have formed due to the mergers and most of the shell features and other substructure in those envelopes will only be visible by using very deep photometry. The \emph{Fornax-analog}s all have shells in the denser part of the stellar distributions as well. These may be very well observable in high quality photometric images. 

All the disks in the $M_{\mathrm{vir}} = 10^{10} \ M_{\sun}\ h^{-1}$ haloes have higher central surface brightness than the Local Group dwarf galaxies (excepting the LMC and SMC) but they are also much more massive. The galaxies closest to the simulated $10^{10} \ M_{\sun}\ h^{-1}$-system are IC4662 and NGC6822 \citep{McConnachie2012}. The \emph{Fornax-analog} has much lower final central surface brightness than the more massive systems, especially initially, and is comparable in surface brightness to the dwarf galaxies in the Local Group in general and the Fornax dwarf galaxy in particular \citep{McConnachie2012}.

\subsubsection{Inner and global shape}
\label{sectshapes}
\begin{figure*}
\centering
\includegraphics[width=0.95\textwidth]{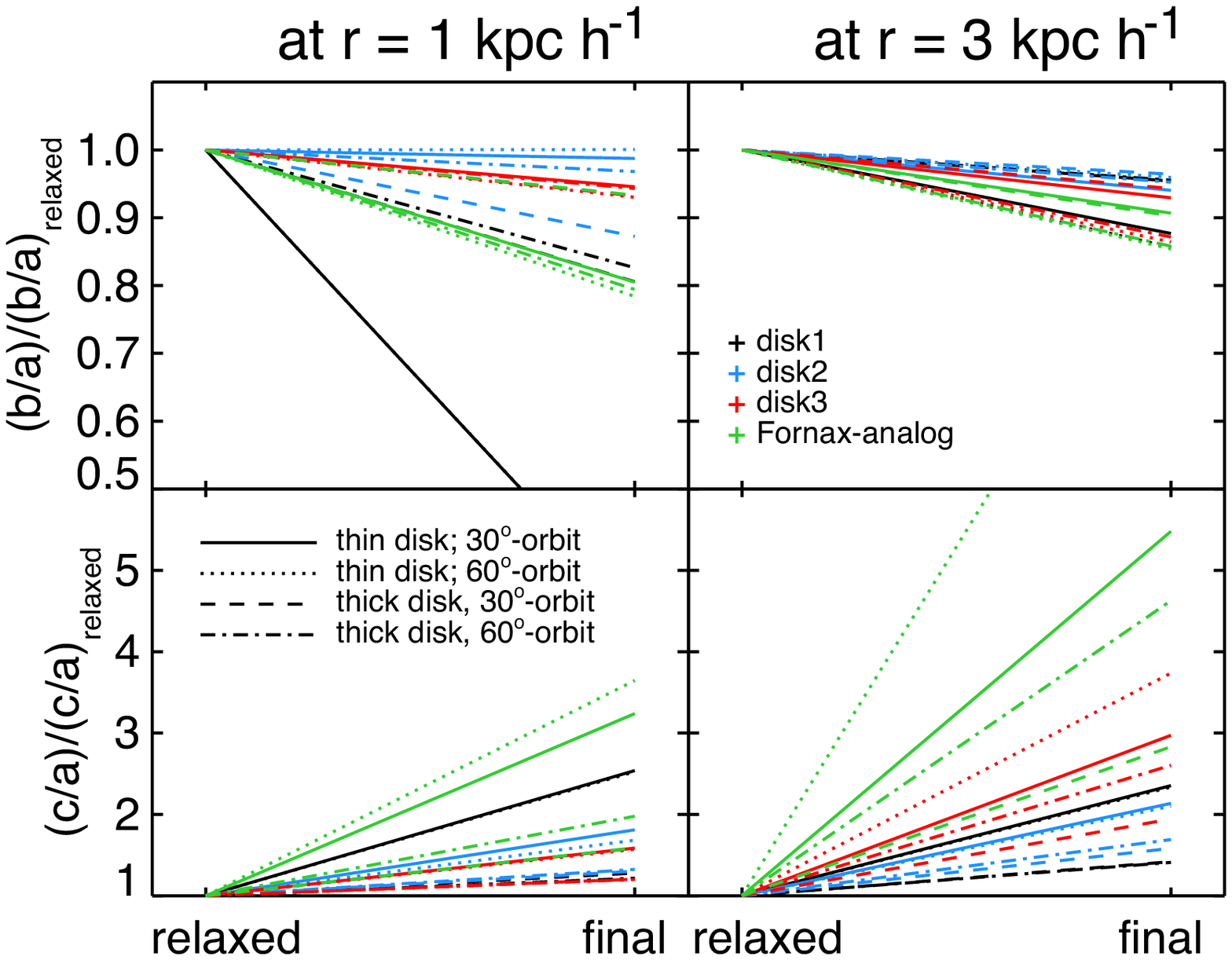}
\caption{\label{shapes} Relative increase in principal axis ratios between initial and final snapshot for all merger simulations. Top: intermediate axis over major axis ($b/a$); bottom: minor axis over major axis ($c/a$). The axis ratios are computed with the major axis length equal to 1 kpc $h^{-1}$ for the panels on the left and equal to 3 kpc $h^{-1}$ for the panels on the right. Shown are 16 simulations of four disky dwarf galaxies with two sets of initial scale heights and two different satellite orbits: the most masive disk (\emph{disk1}; black), the intermediate disk (\emph{disk2}; blue), the least massive disk (\emph{disk3}; red) and the \emph{FNX-analog} (green). (The color and line coding is the same as in Figs. \ref{figorbits} and \ref{figmass}.) Note that the $b/a$- and $c/a$ values are relative to the initial value so they can both become greater than 1.}
\end{figure*}

The difference in the final morphology of the systems can also be seen when fitting an ellipsoid to the stellar mass distribution of the particles in the simulation. We compute the shape tensor 
\begin{equation}
\label{shapetensor}
S_{ij} = \sum_{k} (r_k)_i (r_k)_j
\end{equation}
where $(r_k)_j$ denotes the $j$-component of the position vector of particle $k$. The tensor $S_{ij}$ is often used with a ``normalization'' by the ellipsiodal radius of the particles within the coordinate system of the ellipsoid \citep[see for example][]{Allgood2006} but the effect is debated \citep{Zempetal2011}. We do not normalize and are therefore effectively weighing the outskirts more.  Figure \ref{shapes} shows the change in the intermediate/major ($b/a$; top) and minor/major ($c/a$; bottom) axis ratios for all particles within ellipsoids with major axes of 1 (left) and 3 (right) kpc $h^{-1}$. The smaller ellipsoid thus focuses on the change in shape in the inner parts of the remnants while the panels on the right show the changes in the shape of the total disk (within an ellipsoidal radius of 3 kpc $h^{-1}$, so including the inner parts as well). Note that in Eq. (\ref{shapetensor}) we do not have to take into account the mass of each particle as the mass of all disk particles is the same. 

The left upper panel of Fig. \ref{shapes} shows that most systems have lower $b/a$ axis ratios at the end. This is due to the presence of bars in the case of the most massive disks. The most extreme cases (both thin \emph{disk1}s) have final average  $b/a_{\mathrm{1 kpc}} \sim 0.3$ and the least massive bar of these simulations (\emph{disk1-60-thick}) still has a final average $b/a=0.81$ within an ellipsoidal radius of 1 kpc $h^{-1}$. The isolated \emph{disk1} evolved on a longer timescale also develops a bar. During the merger the bar greatly increases in mass however, and has a major axis of around 1 kpc~$h^{-1}$, extended beyond the half-mass radius of the remnant ($r_{1/2} \sim r_{1/2 \mathrm{;initial}}  \sim 0.8$ kpc $h^{-1}$).  

The bar in \emph{disk1} is also reflected in the final intermediate-over-major axis ratio within an ellipsoid of 3 kpc $h^{-1}$ due to its contribution in the inner parts. Note that although the axis ratios stay close to $b/a\sim 1$ for \emph{disk2} and \emph{disk3} there is also substructure in the inner disks of these systems (see Figs. \ref{SB1} and \ref{SB2}) but not massive enough to define the average shape of the disks. The final $b/a$-ratio at larger radii however, is less axissymmetric for both the \emph{disk3}s with the satellite on the 60-degrees orbit, probably due to the significant tilt of the disk during the merger and the resulting warps.

All \emph{Fornax-analog}s are not axisymmetric anymore in the final snapshot as their intermediate-over-major axis ratio drops to $b/a \sim 0.8$--$0.9$ (depending on location and experiment). Nevertheless, when these relatively small changes in the intermediate-over-major axis ratios are compared to the changes in the minor-to-major axis ratios one notices that the shapes of all the \emph{FNX-analog} systems evolve drastically. These changes in the minor-to-major axis ratios are overall very diverse. The bottom panels of Fig. \ref{shapes} show that some of the \emph{FNX-analog} systems would be classified as spheroidal or elliptical; their total minor-over-major axis ratios grow extremely with factors ranging from almost 3 to almost 10. The initial $c/a \sim 0.06$ (thinner system) and $c/a \sim 0.16$ (thicker system) have resulted after a minor merger in final values in the range $c/a \sim 0.31$--$0.73$. 

Note that the axis ratios plotted in Fig. \ref{shapes} are with respect to the values after relaxation, before the merger, and thus show the heating of the disks are due to the minor merger and not to numerical artifacts. The thickening due to relaxation of the initial disks is very minor. The \emph{disk1}s and \emph{disk2}s have relaxed axis ratios of $c/a \sim 0.075$ for the thin disks and $c/a \sim 0.115$ for the thick disks instead of the initial $c/a \sim 0.05$ (thin) and $c/a \sim 0.10$ (thick). As a second check we have simulated the disks without a merger over the same complete time period, and found no significant changes. As explained earlier, only \emph{disk1} develops a massive bar in isolation.
 
Both the initially thinner and thicker \emph{FNX-analog} systems merging with the satellite on the 60-degrees orbit have $c/a > 0.5$ while also having lower $b/a$-values: $c/a_{\mathrm{FNX-60-thin}} \sim 0.55$ compared to $b/a_{\mathrm{FNX-60-thin}} \sim 0.85$ and $c/a_{\mathrm{FNX-60-thick}} \sim 0.73$ compared to $b/a_{\mathrm{FNX-60-thick}} \sim 0.85$. The \emph{FNX-analog} systems with the satellite on a 30-degrees orbit have lower final global minor-over-major axis ratios, $c/a \sim 0.31$ for the initial thinner and $c/a=0.45$ for the initial thicker system, and are slightly triaxial with $b/a \sim 0.90$. This means that this object viewed from a random angle with have an observational axis ratio that is greater or equal to $0.3$.

For all systems, \emph{disk1}, \emph{disk2}, \emph{disk3} and \emph{FNX-analog}, the thinner disk experiences more thickening in a relative (but not absolute) sense than those that were thick initially. This holds for the final/initial minor-over-major axis ratios for both just the inner regions and for the whole remnant. For \emph{disk1}, especially the initially thinner disks, the increase in disk height is probably dominated by the formation of the substantial bars. Otherwise $c/a$ increases more for the less massive disks than for those more massive; the increase in $c/a$ is inversely ranked by mass. This reflects the fact that a less massive disk is more prone to significant perturbations by the incoming satellite and so dwarf galaxies with smaller baryon fractions are more easily disturbed and will more often have thickened or perturbed morphologies (as also shown in Figs. \ref{figmorph1} to \ref{SB2}).

\subsubsection{Radial and vertical density profiles}
\label{sectprof}
\begin{figure*}
\centering
\includegraphics[width=0.9\textwidth]{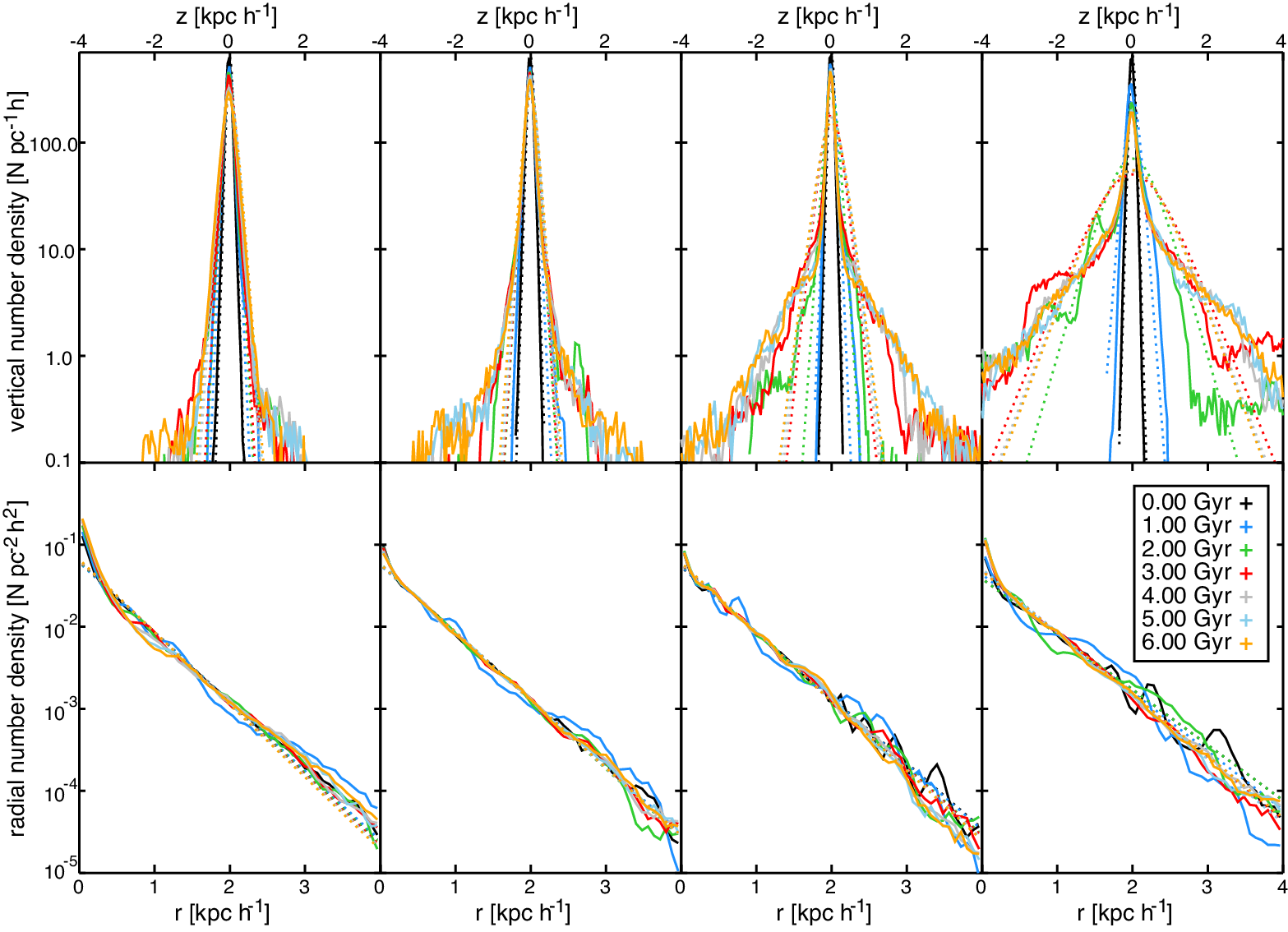}
\caption{\label{densprof} Evolution of the vertical and radial density profiles for the different thin disks ($z_0 = 0.1 R_d$ initially) with the satellite on the 60-degrees orbit: \emph{disk1} (left), \emph{disk2} (middle left), \emph{disk3} (middle right) and \emph{FNX-analog} (right). Density profiles (solid lines) are shown for the initial relaxed disk (0 Gyr) up to 6 Gyr later (about 5 Gyr after the first pericentric passage of the satellite) in steps of 1 Gyr. The dotted lines in the top row show the maximum likelihood $\mathrm{sech}^2$-profiles, in the bottom row the same for an exponential profile, taking into account all disk particles.}
\end{figure*}

In the top panels of Fig. \ref{densprof} we plot the vertical density profiles of the \emph{60-thin} simulations for all systems. Note that we define the plane of the disk to be perpendicular to the total angular momentum vector of all stellar particles within the half mass radius of the stellar remnant. The vertical density profiles of the remnant systems are not well described by the initial $\mathrm{sech}^2$-function. This can be seen through a comparison to the dotted lines which correspond to a maximum likelihood fit to a $\mathrm{sech}^2$-profile. 

The top-right panel shows the density profile of the initially thinner \emph{FNX-analog} system during the merger simulation with the satellite on the orbit with an inclination of 60 degrees. Already after 1 Gyr there are significant deviations from a $\mathrm{sech}^2$-profile. The second panel in the top row shows that for \emph{disk2-thin-60} the vertical density profiles broadens but can still be approximately described by a $\mathrm{sech}^2$-function even close to the time of the merger itself (around 1--2 Gyr) and at the end of the simulation (6 Gyr). The top-left panel shows that for \emph{disk1} the disk thickens more after the actual merger and that the changes in the vertical density profile are strong. This is due to the formation of a massive bar. The \emph{disk3} (top row, middle right panel) has a significant amount of mass at large radii that corresponds to the extended envelope seen in Fig. \ref{figmorph2} in all \emph{disk3} simulations after first pericenter and this renders a much worse $\mathrm{sech}^2$-fit. For \emph{disk2} and \emph{disk1} this is much less the case although as stated earlier a $\mathrm{sech}^2$ functional form does not really reproduce the vertical structure of our remnants.

The progressively worse fits to the $\mathrm{sech}^2$-function, going from \emph{disk1} to \emph{FNX-analog} are due to the increased importance of extra-planar material, which is manifested in extended wings. One may argue that just like in the case of Milky Way-like simulations discussed in the literature, a second thicker component has formed as a result of the merger.

The bottom panels of Fig. \ref{densprof} show the radial surface density profiles with time in steps of 1 Gyr. Note the significant disturbance in all disks around the time of first pericenter (blue lines). The lower mass disk have some spiral structure initially and fall of more steeply than exponential profiles in the outskirts at later times. The most massive disk (\emph{disk1}; leftmost panel in Fig. \ref{densprof}) on the other hand shows an growing overdensity in the very center which corresponds to the massive bar forming in the inner parts of these remnants.  Note that even the remnant of the \emph{FNX-analog} can still be reasonably well fitted by a radial exponential profile as indicated by the dotted lines.

\subsubsection{Comparison to other work on disk thickening}

We compare our results of the thickening of disks due to minor mergers with a number of results from the literature. To this end we measure the thickness of our systems between 2 and 3 disk scale lengths (what would correspond to the solar neighbourhood in the case of a Milky Way-like simulation). We compute two commonly used estimates of the thickness of a disk: 
 the standard deviation of the vertical distribution of particles, $\langle z^2\rangle^{1/2}$ \citep{VelazquezWhite1999}, and a Bayesian maximum likelihood estimate assuming a $\mathrm{sech}^2$-profile.
 
If the vertical density profiles are well described by the function $\frac{1}{\pi z_0} \ \mathrm{sech}^2\left(\frac{z}{z_0}\right)$ then  $\langle z^2\rangle^{1/2} = \frac{\pi}{2 \sqrt{3}} z_0$. Therefore a comparison of these two different measures gives an indication in how far the disk still follows a $\mathrm{sech}^2$. The results can be found in Table \ref{z0}. Here we list the initial values for the scale heights of the disks (after relaxation in isolation) as well as those after the merger. Note that, for \emph{disk1} and \emph{disk2} the thinner disks depict a larger increase in scale height than their corresponding thicker disks (although in an absolute sense the latter are thicker at the final time), and that the scale heights of all \emph{disk1}s increase significantly due to massive bars. For the lower mass disks the mean scale height increases most in simulations with the satellite on the 60-degrees orbit.

  \begin{table}
  \caption[]{\label{z0} Estimates for the scale height $z_0$}
          $$ 
         \begin{array}{llrr}
            \hline
            \noalign{\smallskip}
            \textrm{System}  &  \textrm{time and} & z_0 \textrm{ from} & z_0 \textrm{ from} \\
             &  \textrm{satellite} & \textrm{max-likelihood} & \qquad \langle z^2\rangle^{1/2} \\
             &  \textrm{orbit} & [\mathrm{kpc}\  h^{-1}] & [\mathrm{kpc}\ h^{-1}] \\ 
          \noalign{\smallskip}
          \hline
          \noalign{\smallskip}
          \mathrm{\emph{disk1-thin }} & \mathrm{ relaxed}   & 0.08 & 0.08 \\
            & \mathrm{ final 30}  & 0.20 & 0.20 \\
            & \mathrm{ final 60}  & 0.18 & 0.19 \\
              \hline
         \mathrm{\emph{disk1-thick }}  & \mathrm{ relaxed}   & 0.12 & 0.12 \\
          & \mathrm{ final 30}  & 0.17 & 0.17 \\  
          & \mathrm{ final 60}  & 0.16 & 0.16 \\
         \hline
         \mathrm{\emph{disk2-thin }} & \mathrm{ relaxed}    & 0.08 & 0.08 \\
          & \mathrm{ final 30}   & 0.15 & 0.15 \\
          & \mathrm{ final 60}   & 0.14 & 0.14 \\
         \hline
         \mathrm{\emph{disk2-thick }} & \mathrm{ relaxed}   & 0.12 & 0.12 \\
          & \mathrm{ final 30}  & 0.19 & 0.19 \\
          & \mathrm{ final 60}  & 0.17 & 0.17 \\
         \hline
         \mathrm{\emph{disk3-thin }} & \mathrm{ relaxed}    & 0.06 & 0.06 \\
          & \mathrm{ final 30}   & 0.14 & 0.15 \\
          & \mathrm{ final 60}   & 0.23 & 0.46 \\
         \hline
         \mathrm{\emph{disk3-thick }} & \mathrm{ relaxed}   & 0.11 & 0.11 \\
          & \mathrm{ final 30}  & 0.17 & 0.18 \\
          & \mathrm{ final 60}  & 0.28 & 0.49 \\
         \hline
       \mathrm{\emph{FNX-thin }} & \mathrm{ relaxed}   & 0.08 & 0.08 \\
         & \mathrm{ final 30} & 0.65 & 0.86 \\
         & \mathrm{ final 60} & 1.33 & 1.37 \\
            \hline    
       \mathrm{\emph{FNX-thick }} & \mathrm{ relaxed}  & 0.20 & 0.20 \\
         & \mathrm{ final 30} & 0.92 & 1.10 \\
        & \mathrm{ final 60} & 1.25 & 1.26 \\
      \hline
     \end{array}
   $$ 
\tablefoot{The scale height $z_0$ is estimated assuming a $\mathrm{sech}^2$-function (see Eq. (\ref{diskdensprof})) for the vertical density profile.}
   \end{table}

\begin{figure*}
\centering
\includegraphics[width=0.95\textwidth]{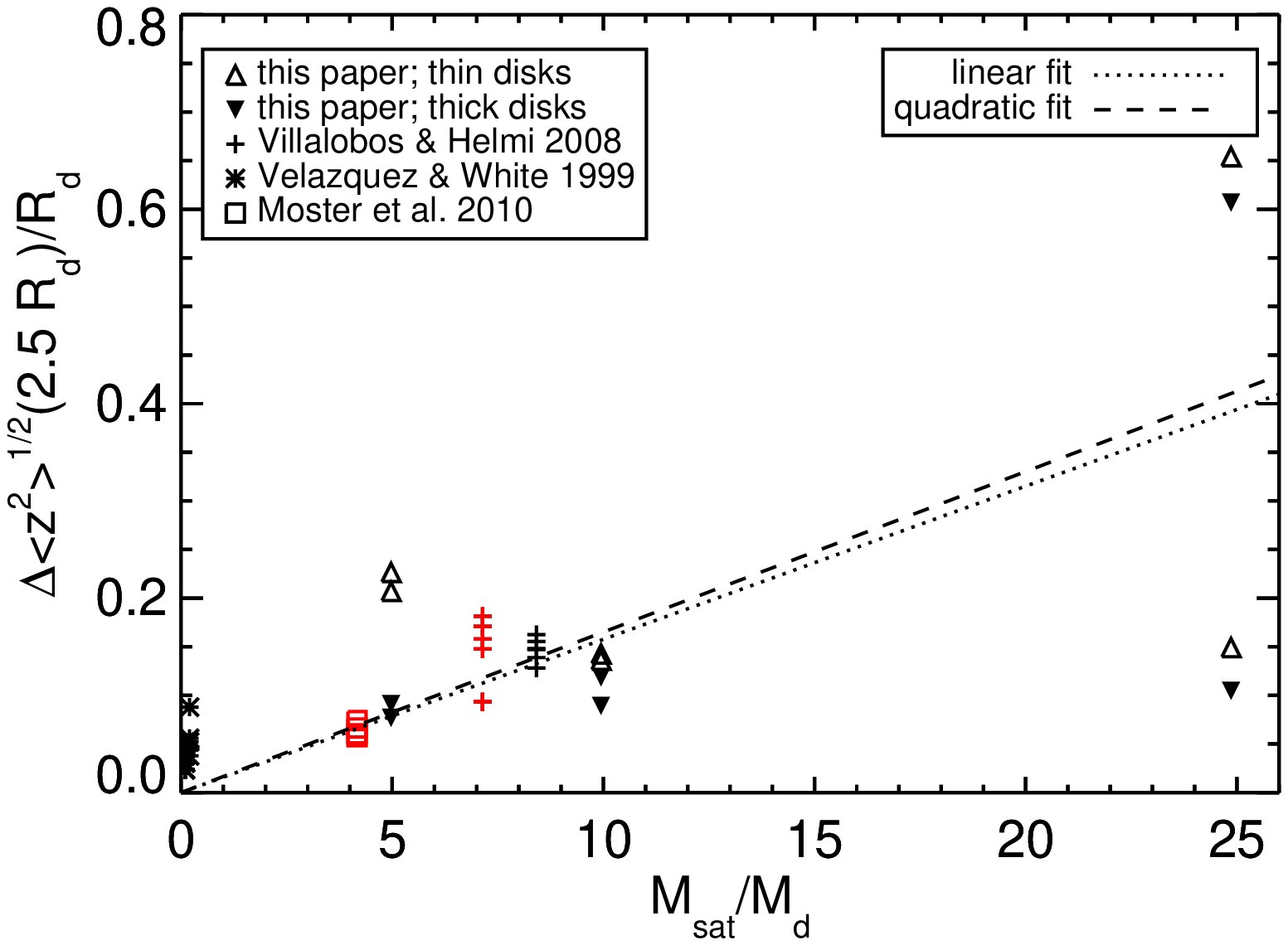}
\caption{\label{Hplot} The change in vertical second order moment of all star particles located near $2.5 R_d$ (the solar radius in a Milky Way-like host), normalized by disk scale length, versus the ratio of satellite to host disk mass for all simulations presented here. In addition we have included results from the literature for similar studies of minor mergers. The \emph{FNX-analog} simulations are not presented because they fall above the vertical range of this plot. The red symbols indicate that the values are read off from a plot instead of taken from a table of the paper in question. Triangles pointing upward denote initially thinner disks while the initially thicker disk are denoted by downward pointing triangles. The dotted and dashed lines show a linear and quadratic fit respectively to the mean values of all $M_{\mathrm{sat}}/M_{\mathrm{disk}}$-bins.}
\end{figure*}

Figure \ref{Hplot} shows the ratio of the difference in the vertical second moment $\Delta \langle z^2\rangle^{1/2}$ to the initial scale length of the disk versus the ratio of the total mass of the satellite and the mass of the disk. In this figure we have added results from minor merger simulations from the literature, specifically from \citet{VelazquezWhite1999}, \citet{VH08} and \citet{Mosteretal2010}. To uniformize the information provided by all these studies, we derive $\Delta \langle z^2\rangle^{1/2}$  assuming that the vertical density distribution in the simulations from \citet{VH08} and \citet{Mosteretal2010} is well represented by a $\mathrm{sech}^2$-function. Note that the four \emph{FNX-analog}-disks have such an increase in $\langle z^2 \rangle^{1/2}$ that they fall completely outside the plot range (see Table \ref{z0}). 

The heating in disk galaxies caused by minor mergers has been proposed to depend linearly \citep{TothOstriker1992, MoVandenBoschWhite2010} or quadratically \citep{Hopkinsetal2008} on the ratio $M_{\mathrm{sat}}/M_{\mathrm{disk}}$. With the exception of this paper all studies so far modeled Milky Way-like disk galaxies disturbed by satellites that are small or at best comparable with respect to the disk. The mergers discussed here however have $M_{\mathrm{sat}}/M_{\mathrm{DM;host}}$ constant and vary $M_{\mathrm{disk}}$, with $M_{\mathrm{disk}}<<M_{\mathrm{sat}}$. 

We apply a least squares fit to all data points in Figure \ref{Hplot}  for a quadratic (with the coefficient of the quadratic part forced to be positive as predicted by \citet{Hopkinsetal2008}) and a linear dependence on $M_{\mathrm{sat}}/M_{\mathrm{disk}}$. We use the mean values of the points per mass bin and exclude the extreme points of the \emph{FNX-analog}s. For our simulations the relation now solely depends on $M_{\mathrm{sat}}/M_{\mathrm{disk}}$. As can be seen from Fig. \ref{Hplot} the scatter is very large, the obtained fits are quite similar, and unsurprisingly neither very good. The reason for this is that the heating of stellar disks by infalling satellites does not solely depend on their mass ratios. For example, the initial structure of the disk and the presence of gas in the disk as well as the orbit of the satellite have a very significant influence on the response of the disk on the infalling satellite \citep[e.g.][]{Mosteretal2010}. Moreover the measurement of heating in different studies is done using different methods and assumptions. Although we have attempted to homogenize this as far as possible further scatter is introduced (e.g. in the case of the \emph{FNX-analog}s the disk is fully destroyed and the characterization by a $\mathrm{sech}^2$ is very poor). From Fig. \ref{Hplot} it is clear that these uncertainties dominate for the larger $M_{\mathrm{sat}}/M_{\mathrm{disk}}$-ratios. 

The parameters of the linear term are almost the same for both the linear and quadratic dependence on $M_{\mathrm{sat}}/M_{\mathrm{disk}}$ (with the coefficient of the quadratic term six orders of magnitude smaller than the coefficient of the linear term: $y=0.016 x$ versus $y=0.017x+9.1\mathrm{e}^{-10}x^2$). Although this study is limited in number, and the scatter in the relation is very significant, it suggests that to first order the heating of galactic disks by minor mergers as a function of $M_{\mathrm{sat}}/M_{\mathrm{disk}}$ can be sufficiently described with a linear model.

   \begin{figure}
   \includegraphics[width=.45\textwidth]{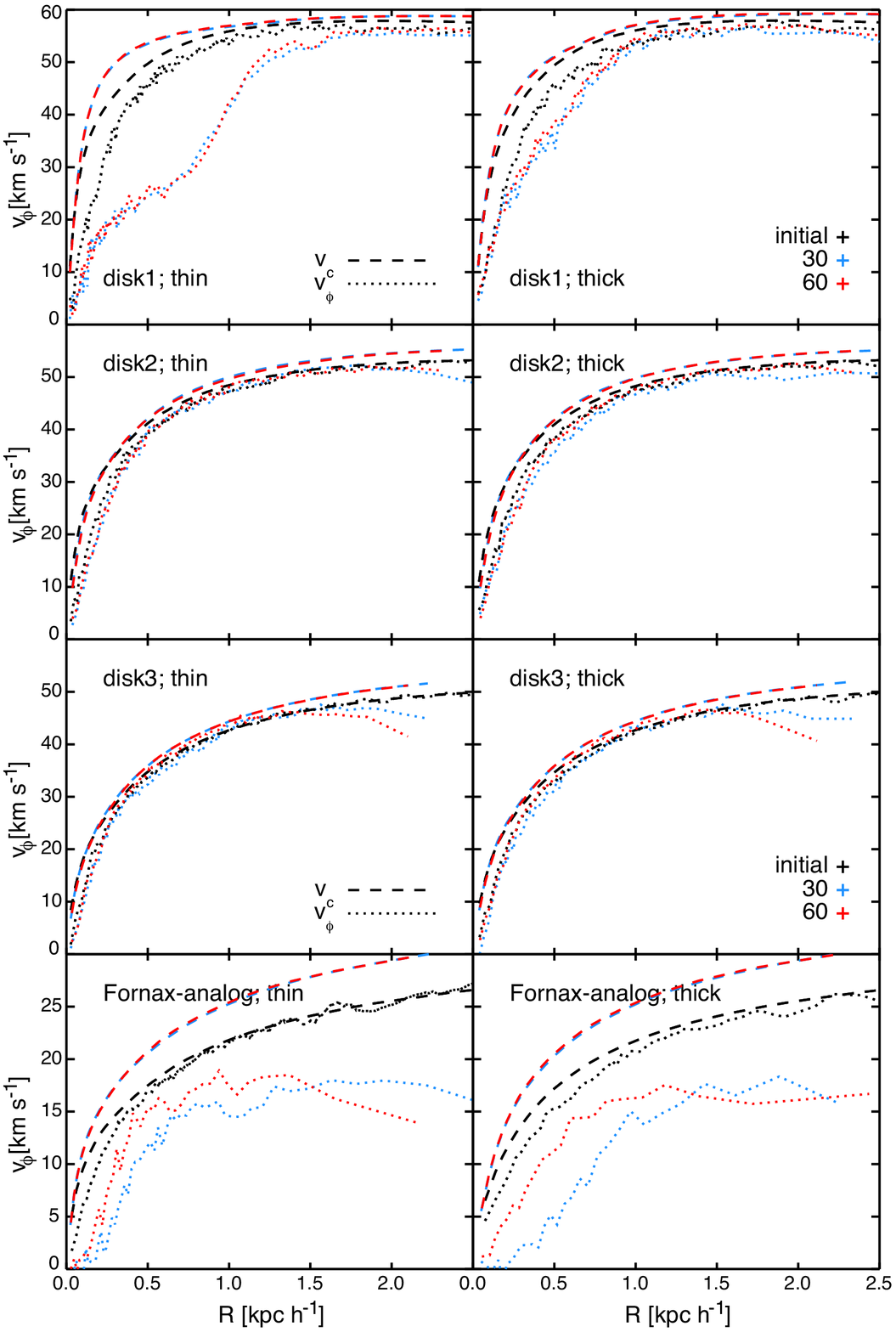}
   \caption{\label{vphi} Circular velocity (dashed lines) and average rotation (dotted lines) of all the disks before the merger (black) and 6 Gyr later for both orbits (30-degrees inclination (blue), and 60-degrees inclination (red)). In all cases the disks are first aligned such that the total angular momentum vector of all particles within the halfmass radius of the disk is perpendicular to the plane in which the average rotational velocity is measured. All particles considered have $|z| < 0.05$ kpc $h^{-1}$ and the bins have a variable binsize but a fixed number of 400 particles each.}
   \end{figure}
   
   \begin{figure}
   \includegraphics[width=.45\textwidth]{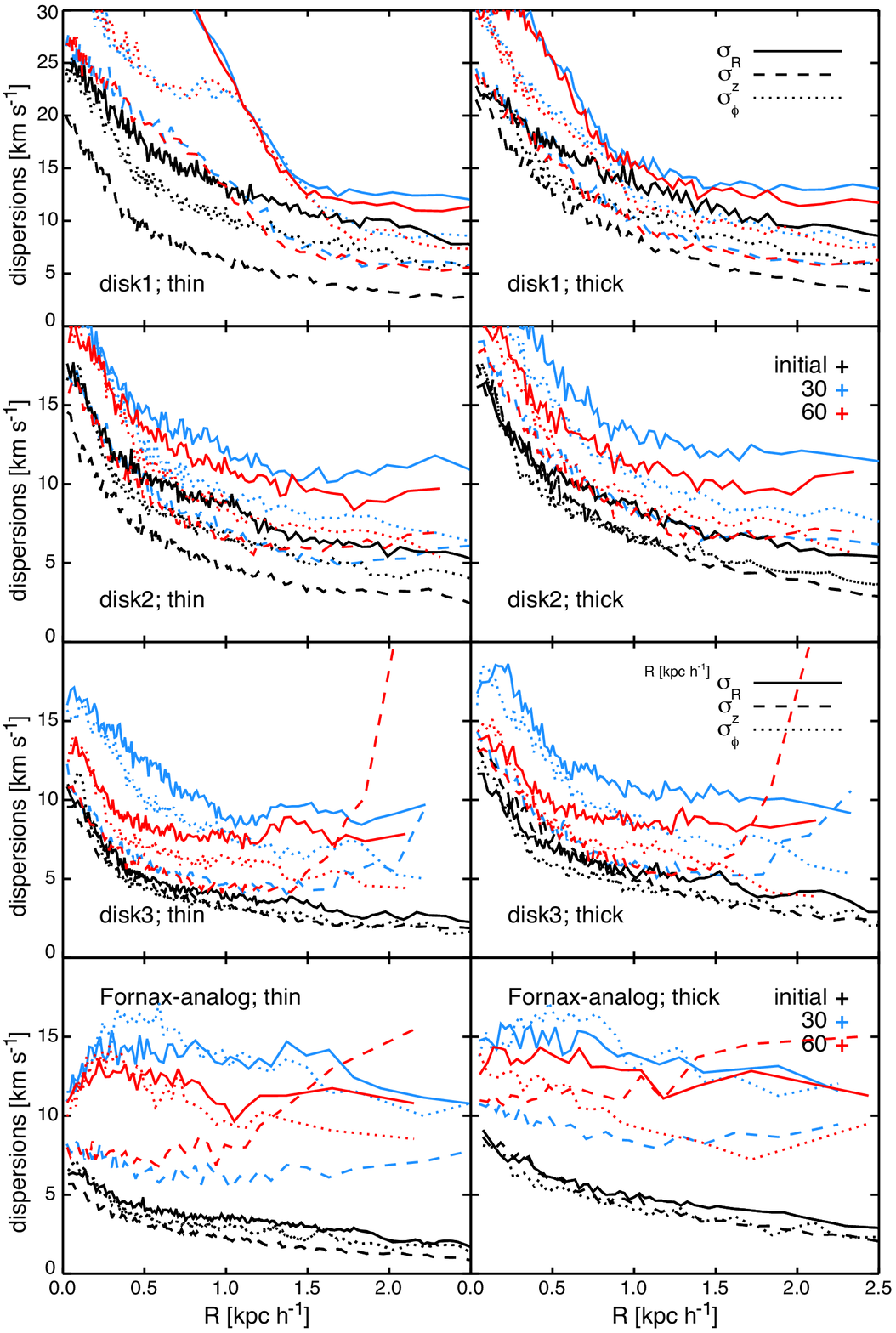}
   \caption{\label{vdisp} Radial (solid lines), vertical (dashed lines) and azimuthal (dotted lines) velocity dispersions of all the disks before the merger (black) and 6 Gyr later for both orbits (30-degrees inclination (blue) and 60-degrees inclination (red)). In all cases the disks are first aligned such that the total angular momentum vector of all particles within the halfmass radius of the disk is perpendicular to the plane in which the velocity dipsersions are measured. All particles considered have $|z| < 0.05$ kpc $h^{-1}$ and the bins have a variable binsize but a fixed number of 400 particles each.}
    \end{figure}

\subsection{Kinematical changes}
\label{kinematics}
The changes induced by the merger events on the rotational velocity and velocity dispersions of the host galaxies can be seen in Figs. \ref{vphi} and \ref{vdisp}. Note that in all cases the circular velocity has increased due to the mass deposited by the accreted satellite in the inner parts of the halo. For all the \emph{disk1}s additional changes are mostly due to the massive bar that forms in these systems. The bar leads to the characteristic linear $v_{\phi} (R)$-relation of solid body rotation, and a steepening of the circular velocity curves for small $R$. For \emph{disk2} and \emph{disk3} the rotational velocities do not change significantly. In the case of \emph{disk3} the rotational velocity drops in the outer parts where the disk is most perturbed. For the \emph{FNX-analog} the evolution is stronger and the rotational velocity has clearly decreased. In the initially thicker disk cases the peak rotation in the region within the stellar half mass radius ($\sim15$ km s$^{-1}$ and $\sim12$ km s$^{-1}$) is very comparable in magnitude to the average line-of-sight velocity dispersions in the same region ($\sim$ 10--14 km s$^{-1}$ with  $\sigma_{\mathrm{l.o.s.}} = \sqrt{\sigma_R^2 + \sigma_z^2+\sigma_{\phi}^2}/\sqrt{3}$). For the initially thinner disks the peak rotation is only a few km s$^{-1}$ higher ($\sim16$ km s$^{-1}$ and $\sim18$ km s$^{-1}$). When observing such a system, and depending on the spatial extend of the data, one could probably conclude there is hardly any rotation present.

All velocity dispersions increase for all the disks. For \emph{disk1} and \emph{disk2} the increase is larger in the inner parts than in the outskirts, especially for \emph{disk1-thin} due to the bars. The \emph{disk3}s on the other hand show a much larger increase in the outer parts which probably reflects the significant number of particles that are in the envelope formed during the merger. Therefore particles at one cylindrical radius now populate a much larger range of velocities. Especially $\sigma_z$, sometimes also considered a measure of the heating in stellar disks, increases dramatically in the outskirts of the disk. Unfortunately, due to the low surface brightness of the outskirts it could be very difficult to measure this signature in dwarf galaxies observationally. 

For the \emph{Fornax-analog}s the velocity dispersions in the inner parts of the disk more than double and in some cases even triple with respect to the values before the merger. Furthermore, the velocity dispersions are more or less constant over the whole disk. Note that, as expected, for both the thin and thick \emph{FNX-analog} the radial and azimuthal dispersions increase more for the 30-degrees merger but that the vertical dispersion increases more in the outer disk for the 60-degrees merger.

The resulting kinematics can be compared to literature values of the dwarfs in the Local Group as most recently summarized in \citet{McConnachie2012}. The Fornax dwarf galaxy has an average line-of-sight velocity dispersion of $11.7$ km s$^{-1}$. We can compare this to $\sigma_{\mathrm{l.o.s.}}$ for the four different \emph{FNX-analog} systems presented in this paper. We compute $\sigma_{\mathrm{l.o.s.}}$ averaged within 1.4 kpc, which is the half-mass radius of the \emph{FNX-analog} remnants, and about 2 times the half-light radius of the Fornax dwarf galaxy, and find values in the range $\sim$ 10--14 km s$^{-1}$. These values are higher than similarly calculated line-of-sight velocity dispersions for the \emph{disk3}s ($\sigma_{\mathrm{l.o.s.}} \sim$ 7--12 km s$^{-1}$) while the rotational velocity of \emph{disk3} is more than twice that of \emph{FNX-analog}. For the \emph{Fornax-analog}s $v_{\phi} \sim \sigma_{\mathrm{l.o.s.}}$ around the half mass radius and $v_{\phi} < \sigma_{\mathrm{l.o.s.}}$ for smaller $R$. Therefore we can say that an initially disky dwarf galaxy similar to the \emph{Fornax-analog} systems presented here, can kinematically as well as morphologically be transformed by a minor merger with its own dark satellite to a more spheroidal like system almost akin to those observed. 

\section{Discussion}
\label{Discussion}
We have presented a series of simulations of minor mergers of dwarf galaxies and their own satellites. The satellites are presumed to be completely dark and we have shown here that for $M_{\mathrm{sat}}/M_{\mathrm{host}} = 0.2$ their effect on the morphology and kinematics of the dwarf galaxies can be severe. Moreover, for larger satellite-to-disk mass ratios the impact of the merger is notably stronger. Therefore it might be essential to take into account such events when considering the evolution of dwarf galaxies. 

When setting up the simulations of the dwarf galaxies in the simulations presented in this paper we were faced with a number of problems. We found that the standard methods used for setting up simulations of Milky Way-like galaxies could not be applied. Especially the velocity structure needed to be revised. As a result of the dominance of the halo potential at all radii, new assumptions have to be made to solve for the velocity structure. In this paper we decided to use the epicyclic approximation as long as it gives physical results and assume that the azimuthal velocity is continuous from that point inwards. This solution is not ideal and results in an azimuthal velocity dispersion that has as unphysical bump. This bump disappears however when relaxing the disks, and the resulting rotation and dispersions are satisfactory. A possible improvement could be to fit a spline, instead of a second order polynomial as we have done.

The concentration of the host halo plays a significant role in how much damage is induced by the satellite, as shown by comparing the \emph{disk3} and \emph{FNX-analog} experiments. A dwarf galaxy with a dark matter density profile that is shallower than seen in dark-matter-only simulations \citep[as perhaps expected because of baryonic effects][]{Mashchenkoetal2008, Governatoetal2010, Governatoetal2012, PontzenGovernato2012, Teyssieretal2013} is therefore much more vulnerable to minor merger events by dark satellites. 

It may be argued that some of the satellites in our simulations (e.g. in the case of \emph{disk1}, \emph{disk2} and \emph{disk3}) are on the verge of being too massive to be dark. However, the virial mass below which haloes are expected to be devoid of stars does not have a sharp transition but is expected to be around 10$^8$--10$^9 \ M_{\sun}$ \citep{Gnedin2000, Hoeftetal2006, Crainetal2007, Okamotoetal2008, GnedinTassisKravtsov09, LiDeLuciaHelmi2010, Sawalaetal2013} \citep[but see also][]{TaylorWebster2005, Warrenetal2007}. If the satellite would have contained a small amount of baryons we might expect a similar result as if the completely dark satellite would be more concentrated (as for the smallest haloes the effects of baryons appear to be small \citep[e.g.][]{Governatoetal2012}). We have tested cases with different dark matter halo concentrations for the dark satellite and in line with our expectations a more concentrated satellite survives longer and therefore has a stronger effect on (or causes more thickening of) the disk of the host dwarf galaxy. 

On the other hand, although we vary the disk mass fraction to lower values (especially \emph{disk3}) as is expected for dwarf galaxies, one could argue that our disks are still too massive for their halo mass. Following the (at this mass extrapolated) stellar mass-halo mass relations from either abundance matching \citep[e.g.][]{Mosteretal2013}, or from hydrodynamical cosmological simulations \citep[e.g.][]{Sawalaetal2014}, our lowest stellar disk mass fraction should be a factor 2--5 smaller. For such a disk the effect of the merger would increase significantly.

An important issue is how often encounters such as those described in this paper will take place. Although the subhalo mass function is almost self-similar and independent of host halo mass, higher satellite-to-disk mass ratios lead to more disruptive events. So the expected fraction for significantly disruptive minor mergers for dwarf galaxies is not only given by the ($M_{\mathrm{sat}}:M_{\mathrm{host}} = 1:5$) fraction but the fraction for minor encounters with $M_{\mathrm{sat}}/M_{\mathrm{disk}}>1$. According to \citet{Deasonetal2014} a significant fraction of the dwarf galaxies in the Local Group (10\% for satellites and 15--20\% for field dwarfs) have experienced major mergers with \textit{stellar} mass ratios $>0.1$. Including dark satellites and broadening the analysis to minor mergers with small pericenters will increase the estimates of dwarf galaxies affected by mergers. \citet{Helmietal2012} show that as the galaxy efficiency or disk mass fraction decreases substantially for lower mass haloes this implies that such an encounter is much more common than previously expected. They estimate that dwarf galaxies similar to \emph{disk3} or \emph{FNX-analog} experience on average 1.5 encounters over a Hubble time where $M_{\mathrm{sat}}/M_{\mathrm{disk}} \simeq 1$ and only taking into account encounters where the pericenter lies within 30\% of the virial radius of the host (dwarf galaxy) halo. A small first pericentric radius ensures a short timescale for the merger and also a large impact. We plan to further quantify the expected significant merger fraction in future work.   

Small galaxies in isolation are generally gas-rich, an element we have not considered in our simulations. Previous work has shown that in mergers the presence of gas influences the response of the stellar component to the tidal forces \citep{Mosteretal2010, Naabetal2006} and the torques that arise during the merger can cause an accumulation of gas close to the center giving rise to an increase in star formation or even a starburst \citep{DiMatteoetal2007, Teyssieretal2010}. We plan to report on simulations of dwarf galaxies including gas in the disk and dark satellites in Starkenburg et al. (in prep.).

\section{Conclusions}
\label{Conclusions}
We have studied the response of disky dwarf galaxies to a $1:5$ minor merger with a dark satellite using a suite of N-body simulations. Our host haloes have masses of $10^{10} \ M_{\sun}\ h^{-1}$ and $4 \times 10^9 \ M_{\sun}\ h^{-1}$ and the baryon fractions range from $0.008$ to $0.04$ assumed to be in a stellar disk. The satellite is completely dark and has a prograde radial orbit with first pericenter within the half mass radius of the disk and an inclination of either 30 or 60 degrees. 

The structure and kinematics of the disks in our simulations are initialized following a carefully adaptation of standard methods to this dwarf scale. We present a method to solve the velocity structure assuming that the potential is dominated by the halo at all radii and only using the epicyclic frequency up to the point where this approximation gives physical results. The resulting velocity profiles after relaxing the disk in isolation are smooth and stable.

In our experiments, the satellite looses mass very quickly while plunging into the denser central regions of the host and is completely disrupted after two to three pericenter passages. All dwarf galaxy disks are significantly disturbed by the minor merger. The most massive disks develop strong bars and thicken due to these bars and stellar envelopes are formed around the disks. Smaller disks form more extended envelopes with a preferential orientations. This can be seen as tidal debris in what would be the inner stellar halo of the dwarf galaxy. However, in our simulations these tidal features are made of material from the disk instead of in the infalling satellite as is often assumed for observed shells, streams and plumes. 

Our suite of simulations shows that the disk mass fraction is an important parameter in minor mergers and that the disk thickens more when the ratio $M_{\mathrm{sat}}/M_{\mathrm{disk}}$ increases. For the lowest mass disks the final remnants\rq{} vertical structure can not be described by a $\mathrm{sech}^2$-function anymore. These systems are spheroidal and their radial profiles follow an exponential form. In these experiments, the azimuthal velocities are comparable in magnitude or even lower than the line-of-sight velocity dispersions. The remnant system has properties similar to the Fornax dwarf spheroidal, indicating that mergers between dwarfs and dark satellites may be an additional channel for the formation of small spheroidal systems.

\begin{acknowledgements}
We are grateful to Laura Sales for many useful discussions and to Carlos Vera-Ciro, Alvaro Villalobos and Volker Springel for providing code. AH acknowledges financial support from
European Research Council under ERC-StG grant GALACTICA-240271.
\end{acknowledgements}

\bibliographystyle{aa} 
\bibliography{TKS} 

\end{document}